\pdfoutput=1
\documentclass[annual]{acmsiggraph}

\usepackage{graphicx}
\usepackage{url}
\usepackage{hyperref}

\usepackage[english]{babel}

\usepackage{amsmath,amsthm,amssymb,bbm}

\theoremstyle{plain}
\newtheorem{theorem}{Theorem}[section] 
\newtheorem{lemma}[theorem]{Lemma} 
\newtheorem{proposition}[theorem]{Proposition} 
\newtheorem*{corollary}{Corollary} 
\theoremstyle{definition}
\newtheorem{definition}{Definition}[section]

\newcommand{\noplus}{}

\newcommand{\tmop}[1]{\ensuremath{\operatorname{#1}}}

\newcommand*\norm[1]{\left\| #1 \right\|}

\title{On the Approximation Theory of Linear Variational Subspace Design}

\author
       {Jianbo Ye\thanks{Email: jxy198@ist.psu.edu}      \\
         College of Information Sciences and Technology\\
         Pennsylvania State University, USA
         \and
         Zhixin Yan\\
         Department of Computer Science\\
         Worcester Polytechnic Institute, USA
}

\TOGonlineid{45678}
\TOGvolume{0}
\TOGnumber{0}
\TOGarticleDOI{1111111.2222222}
\TOGprojectURL{}
\TOGvideoURL{}
\TOGdataURL{}
\TOGcodeURL{}
\pdfauthor{}

\begin{document}


\maketitle

\begin{abstract}
   Solving large-scale optimization on-the-fly is often a difficult task 
   for real-time computer graphics applications. To tackle this challenge, 
   model reduction is a well-adopted technique. Despite its usefulness,
   model reduction often requires a handcrafted subspace that spans
   a domain that hypothetically embodies desirable solutions. 
   For many applications, obtaining such subspaces case-by-case either is impossible or 
   requires extensive human labors, hence does not readily have a scalable solution for 
   growing number of tasks. We propose linear variational subspace design for large-scale constrained 
   quadratic programming, which can be computed automatically without 
   any human interventions. We provide meaningful approximation error bound that substantiates 
   the quality of calculated subspace, and demonstrate its empirical success in
   interactive deformable modeling for triangular and tetrahedral meshes.

\end{abstract}

\begin{CRcatlist}
  \CRcat{I.3.5}{Computer Graphics}{Computational Geometry and Object Modeling}{}
  \CRcat{I.3.6}{Computer Graphics}{Methodology and Techniques}{Interaction techniques}
\end{CRcatlist}

\section{Introduction}

In computer graphics realm, solving optimization with a substantially 
large amount of variables is often an expensive task. 
In order to speed up the computations, model reduction has been 
introduced as a useful technique, particularly for interactive and real-time applications.
In solving a large-scale optimization problem, 
it typically assumes that a desired solution approximately lies 
in a manifold of much lower dimension that is independent 
of the variable size. Therefore, it is possible to cut down calculations 
to a computationally practical level by only exploring variability (i.e., different 
solutions subject to different constraints) in a suitably chosen low-order space,
meanwhile, attempting to produce visually convincing results just-in-time.
In this paper, we re-examine model reduction techniques for quadratic 
optimization with \textit{uncertain} linear constraints, which has been widely used in
interactively modeling deformable surfaces and solids.

Modeling deformable meshes has been an established topic 
in computer graphics for years~\cite{sorkine2004laplacian,yu2004mesh}.
Mesh deformation of high quality is accessible via off-line solving a large-scale optimization
whose variables are in complexity of mesh nodes. A studio work-flow in mesh deformable modeling 
often involves \textit{trial-and-error} loops: an artist tries different sets of constraints and 
explores for desirable poses. In such processes, 
an interactive technique helps to save the computation time 
where approximate solutions are firstly displayed for the purpose of 
guidance before a final solution is calculated and exported. 
Nevertheless, interactive techniques related to real-time mesh modeling has been less successful than
their off-line siblings till today. Existing work based on model reduction 
often requires a high quality subspace as the input,
which typically demands human interventions in constructing them. 
Exemplars include cage-based deformations~\cite{huang2006subspace,ben2009variational}, 
LBS with additional weights~\cite{jacobson2012fast}, LBS with skeletons~\cite{shi2007mesh}, 
and LBS with point/region handles~\cite{au2007handle,sumner2007embedded}. 
The time spent on constructing such reduced models is as much as, if not more than,
that spent on on-site modeling. In industrial deployments, companies have to hire
many artists with expertise skills for rigging a large set of models 
before those models are used in productions. 
This poses the necessity for a fully automatic subspace generation method. This problems
have received attentions in the past. For example, data-driven methods have been developed
for deformable meshes, where a learning algorithm tries to capture the characteristics of 
deformable mesh sequences and applies to a different task~\cite{sumner2005mesh,der2006inverse}. 
However, they still struggle to face two challenges: 1) Scalability:
Like approaches relying on human inputs, obtaining a deformable sequence of scanned meshes can also 
be expensive. No words to say if we want to build a deformable mesh database containing 
large number of models with heterogeneous shapes. 2) Applicability: Many models of complex
geometries or topologies are relatively difficult to rig, and there are no easy ways to build
a set of controllers with skinning weights to produce desirable deformations.
Though we see there have been several workarounds for a domain specific 
mesh sets, such as faces and clothes,
an automatically computed subspace for arbitrary meshes, 
which is cheaply obtained, still can be beneficial, 
if not all, for fast prototyping or exploratory purposes: 
 the set of constraints chosen on-site is exported 
for computing a deformation with full quality in the off-line stage.

In this paper, we introduce an automatic and principled way to create reduced models, 
which might be applied to other computationally intensive optimization scenarios 
other than mesh deformation. Our main idea is very simple: in solving a constrained quadratic programming,
we observe that Karush-Kuhn-Tucker (KKT) condition implicitly defines an effective subspace that can be
directly reused for on-site subsequent optimization. We name this \textit{linear variational subspace} (for short, variational subspace). 
Our contribution is to theoretically study the approximation error bound of variational subspace
and to empirically validate its success in interactive mesh modeling. The deformation 
framework is similar to one used in \cite{jacobson2012fast}.%
\footnote{In independent work reported in a recent preprint~\cite{wang2015linear}, 
Wang et al. also propose a mesh deformation framework based on
linear variational subspace similar to ours with the difference that we in addition
use linear variational subspace to model rotation errors in reduced-ARAP framework.
Our deformation can be similar to theirs~\cite{wang2015linear}, if regularized coefficient $\alpha$ 
is set to a large value. Therefore, the contribution of our paper 
excluding the empirical efforts is the approximation theory
for linear variational subspace.}\footnote{Implementations and demos: \url{https://github.com/bobye}}
We further examine the deformation 
property of our proposed method, and compare with physically based deformation
~\cite{ciarlet2000mathematical,grinspun2003discrete,botsch2006primo,sorkine2007rigid,chao2010simple}
and conformal deformation\cite{paries2007simple,crane2011spin}.

\begin{figure}[tp]
  \includegraphics[width=.45\textwidth]{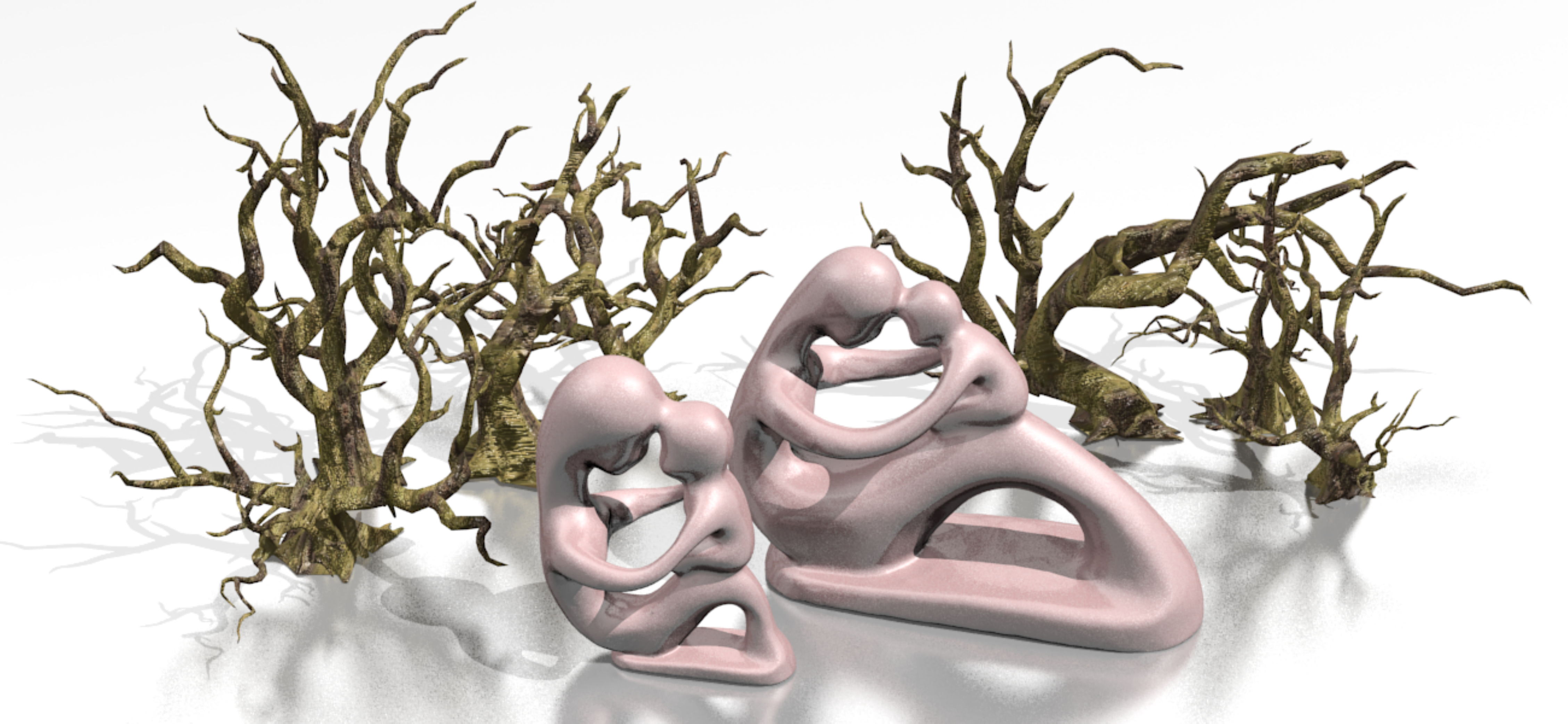}
  \caption{Variational subspace provides robust and high quality deformation results regarding 
    arbitrary constraints at runtime. This figure shows tree and fertility as well as their deformed
    versions. }
  \label{fig:header}
\end{figure}

\section{Mathematical Background}\label{sec:setup}

Consider minimizing a quadratic function  $f (X;q) = (1/2)X^{T} HX-q^{T} X$
subject to linear constraints $A^{T} X=b$,
where $X\in \mathbb R^n$ is an overly high dimensional solution, $H$ is a
semi-definite positive matrix of size $n \times n$, $q\in \mathbb R^n$, 
$A$ is a well-conditioned matrix of size $n \times m$, and $b\in \mathbb R^m$. 
Typically, $m \ll n$. Without loss of generality, we can write
\begin{equation}
  \label{eq:quadprog} \begin{array}{rl}
    \min_{X} & (1/2) X^{T} HX-q^{T} X\\
    \text{s.t.} & A^{T} X=b.
  \end{array}
\end{equation}
Instead of solving the optimization with a single setup, we consider a
set of them with a prescribed fixed $H$, and varying $A$, $b$ and $q$ under certain conditions. 
The ``demand'' of this configuration is defined to be a particular choice of $A$, $b$ and $q$.
Different choices usually result in different optimum solutions. 
When $n$ is relatively small, efficiently solving for unreduced solutions
belongs to the family, so called multi-parametric quadratic programming, 
or mp-QP~{\cite{bemporad2002explicit}}{\cite{tondel2003algorithm}}.
We instead approach to tackle the same setting with a large $n$ by exploring approximate 
solutions in a carefully chosen low-order space. 

We model the ``demand''s by assuming each column of $A_{n\times m}$ is selected
from a low-order linear space $C_{n \times d}$, 
namely $A=C_{n \times d}A_{c}\in \tmop{Span}(C)$ for some $A_c$, 
and $q$ is again selected from another low-order linear space $\tmop{Span}(D)$ such that
that $q=D_{n \times k} Y$ for some $Y$, where $A_{c}$ is a matrix of size $d \times m$,
$Y$ is a vector of size $k \times 1$. Here $d$ and $k$ is the dimension of reduced subspace 
$C$ and $D$ articulating to what $A$ and $q$ belong, respectively. 
Instead of pursuing a direct reduction in domain of solution $X$,
we analyze the reducibility of ``demand'' parameters $A$ and $q$ 
by constructing reduced space $C$ and $D$. 
Specifications of the on-site parameters $A_{c}$, $b$ and $Y$
turn out to be the realization of ``demand''s. We can rewrite {\eqref{eq:quadprog}} as
\begin{equation}
  \label{eq:2stage} \begin{array}{rl}
    \min_{X,Z} & (1/2) X^{T} HX-Y^{T} D_{n \times k}^{T} X\\
    \text{s.t.} & C_{n \times d}^{T} X=Z, \hspace{1em} A_{c}^{T} Z=b.
  \end{array}
\end{equation}
Optimization {\eqref{eq:2stage}} can be decomposed into an equivalent two-stage formulation, i.e.,
\begin{equation}
  \label{eq:2stage2} \min  _{A_{c}^{T} Z=b} \{ \min  _{C^{T} X=Z} f(X;DY) \} .
\end{equation}
Karush-Kuhn-Tucker (KKT) condition yields that the optimum point $X^{\ast}
(Z,Y;C,D)$ for $\min  _{C^{T} X=Z} f (X;DY)$ should satisfy linear equations
\begin{equation}
  \left(\begin{array}{cc}
    H & C\\
    C^{T} & 0
  \end{array}\right) \left(\begin{array}{c}
    X^{\ast}\\
    \Lambda
  \end{array}\right) = \left(\begin{array}{c}
    DY\\
    Z
  \end{array}\right) . \label{eq:qpform}
\end{equation}
where $\Lambda$ is a Lagrange multiplier. Therefore $X^{\ast} (Z,Y;C,D)$ is
affine in terms of on-site parameters $Z$ and $Y$, i.e.,
\begin{equation}
  \label{eq:qpsubspace} X^{\ast} (Z,Y;C,D) =N(;C)Z+U(;C)DY \hspace{0.27em} ,
\end{equation}
where $N(;C)$ and $U(;C)D$ can be computed  before $A_{c}$, $b$
and $Y$ are observed in solving the second stage of {\eqref{eq:2stage2}}:
From Eq.~\eqref{eq:qpform}, each column of $N(;C)$ is computed through 
a preconditioned linear direct solver by setting $Z$ 
as the corresponding column of $I_{d \times d}$ with $Y=0$;
And similarly, each column of $U(;C)D$ is linearly solved by setting
$DY$ to be the corresponding column of $D_{n \times k}$ with $Z=0$.
Remark it is particularly required that $C$ and $D$ has to be 
full-rank and well-conditioned (as will be specified later).
\begin{figure}[t]
    \includegraphics[width=.45\textwidth]{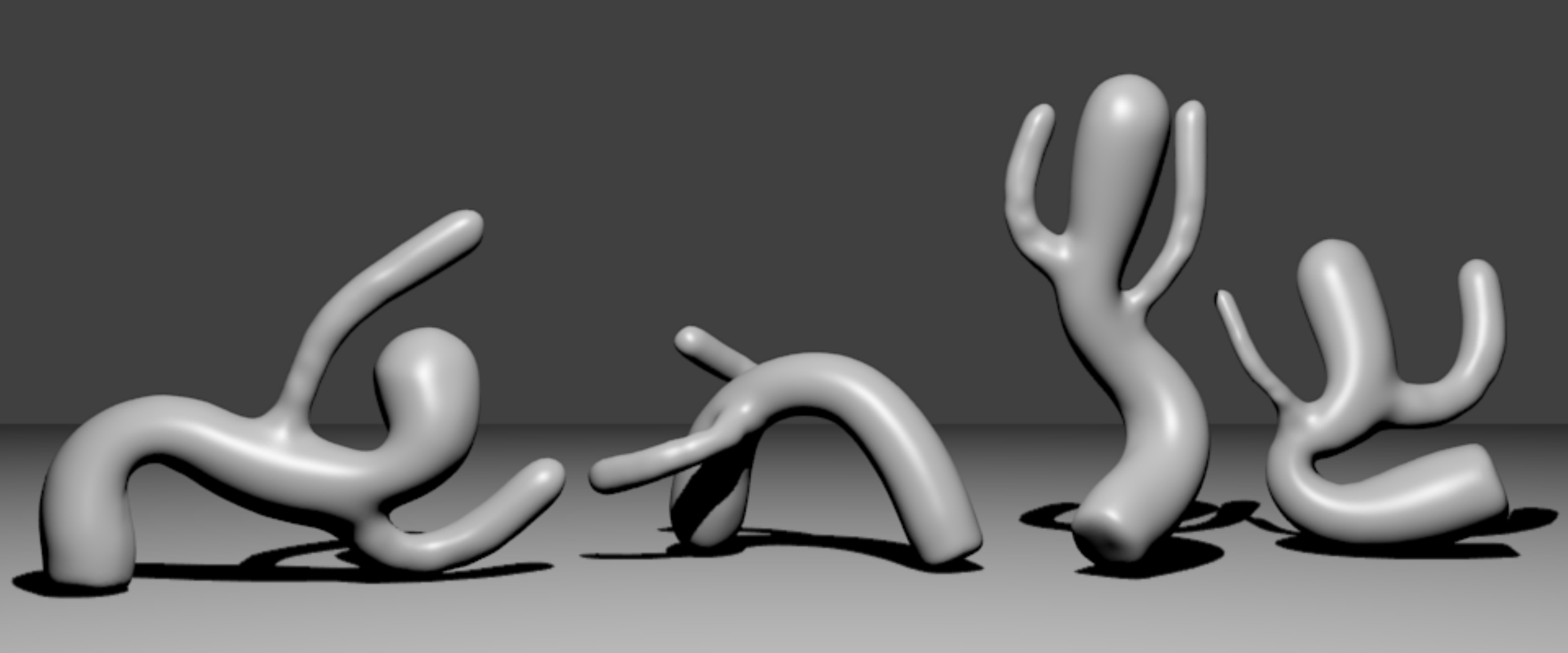}
    \caption{Artistic freedom is important in deformable mesh modeling
      because many artists are interested in authoring creative editing.
      Rig or cage based deformation lacks the richness of deformable variants.}
    \label{fig:cactus}
\end{figure}

Once we have a subspace design {\eqref{eq:qpsubspace}}, for arbitrary ``demand'' $A$, $b$
and $q$, we can immediately solve for an approximate solution $X^*(Z_{\min}, Y_{\min})$ 
via substituting {\eqref{eq:qpsubspace}} into the original formulation {\eqref{eq:quadprog}}.
We clarify that if our assumption is held, i.e., $A=CA_{c}$ and $q=DY$ for some $A_{c}$ and $Y$, 
the approximate solution is identical to the exact optimal $X_{\min}$. 
Furthermore, the approximated solution can work with hard constraints in
the online solvers. 

In next, we demonstrate its effectiveness by implementing an 
interactive mesh deformation method based on the model reduction framework we proposed.
In the end, we will return to the theoretical aspects, and 
derive an error bound for the approximate solution 
with respect to the use of $C$ and $D$.

\section{Interactive Mesh Deformation} \label{sec:ourapproach}
In this section, we start from the point that is familiar to the graphics audiences, and proceed to
the practice of our reduced model, where we mainly focus on deriving
the correct formulation for employing variational subspace. 
Experimental results are provided in the end.
In order for practitioners to reproduce our framework, 
we describe the details of our implementation in Appendix~\ref{sec:implementation}.
We remark that, subspace techniques described in this section
has been standardized as described in~\cite{jacobson2012fast}. The main difference
is to replace the original linear subspace of a skinning mesh with 
the variational subspace described in our paper. Our variational
subspace techniques extends the fast deformable framework 
as proposed in~\cite{jacobson2012fast} to meshes whose skinning
is not available or impossible, such as those of complex typologies. 
There are, however, good reasons to work with linear blending skinning,
for example, it is often possible for artists to directly edit 
the weights painted on a skinned mesh. 

\noindent\textbf{Notations.}
Denote by $\mathbf v_1, \ldots, \mathbf v_n \in \mathbb R^3$
the rest-pose vertex positions of input mesh $\mathcal M$,
and denote the deformed vertex positions by $\mathbf v'_1, \ldots, \mathbf v'_n \in \mathbb R^3$.

Use bold lowercase letters to denote single vertex $\mathbf v\in\mathbb R^3$ and
$3\times 3$ transformation matrix $\mathbf r$, and bold uppercase letters
$\mathbf V$ and $\mathbf R$ to denote arrays of them. We use uppercase normal font letters
to denote general matrices and vectors (one column matrices) and lowercase normal
letters to denote scalars. {We may or may not specify the dimensions of matrices
explicitly in the subscripts, hence $M_{n\times n}$ and $M$ are the same.
For some other cases, subscripts are enumerators or instance indicators. 
We use superscripts with braces
for enumerators for matrices, e.g., $M^{(i)}$.}

Use $\norm{\cdot}$ to denote Frobenius norm of matrices (vectors), 
$\norm{\cdot}_2$ to denote $L_2$ norm of matrices, and
 $\norm{Z}_M = \sqrt{\mbox{tr}(Z^TMZ)}$ to denote Mahalanobis norm 
with semi-positive definite matrix $M$.

Denote dot product of matrices by $\circ$, and Kronecker product of 
matrices by $\otimes$. Let $I$ be the identity matrix, $\mathbf 0$ be the
zero matrix and $\mathbf 1$ be a matrix of all ones.

\subsection{Variational Reduced Deformable Model}\label{sec:algorithm}
\noindent\textbf{ARAP energy.} In recent development of nonlinear deformation energy,
the As-Rigid-As-Possible energy~\cite{sorkine2007rigid,xu2007gradient,chao2010simple}
is welcomed in many related works, in which they represent deformations by
local frame transformation. The objective energy function under this representation is 
quadratic in terms of variables: vertices and
 transformation matrices with orthogonality constraints. 
This family of energy functions can be written as
\begin{equation}
  \mathcal E(\mathbf V', \mathbf R) =  \frac 12 \sum\limits_{k=1}^r \sum\limits_{(i,j)\in \mathcal G_k}\!\!\! c_{ijk} \norm{(\mathbf v'_i\!-\!\mathbf v'_j)\!-\!\mathbf r_k (\mathbf v_i\!-\!\mathbf v_j)}^2\;,
\end{equation}
where $\mathcal G_k$ are their corresponding
sets of edges (see Figure 5 of~\cite{jacobson2012fast}),
$c_{ijk}\in \mathbb R$ are typically the cotangent weights~\cite{chao2010simple},
and $\mathbf r_k\in SO(3)$ denotes the local frame rotations.
By separating quadratic terms and linear terms w.r.t. $\mathbf v_i$,
and vectorizing $(\mathbf v_i)_{i=1}^n$ and $(\mathbf r_k)_{k=1}^r$ to
$V'_{3n\times 1}$ and $R_{9r\times 1}$ respectively,
ARAP energy can be further expressed as
\begin{equation}\label{eq:arapenergy}
  E(V',R) = \frac 12 V'^T H V' -  R^T K V' + \mbox{constant},
\end{equation}
where $\mathbf r_k^T \mathbf r_k = I_{3\times 3}$ (see \cite{jacobson2012fast} for more details).\\

\noindent\textbf{Rotational proxies.}
By observation, minimizing ARAP energy involves solving $R$,
which is in complexity of mesh geometries.
We modify the original ARAP energy to a piece-wise linear form,
which relieves the high non-linearity of optimization, but simultaneously
increases the complexity by introducing linearization variables.

\begin{figure}[t]
    \includegraphics[width=0.45\textwidth]{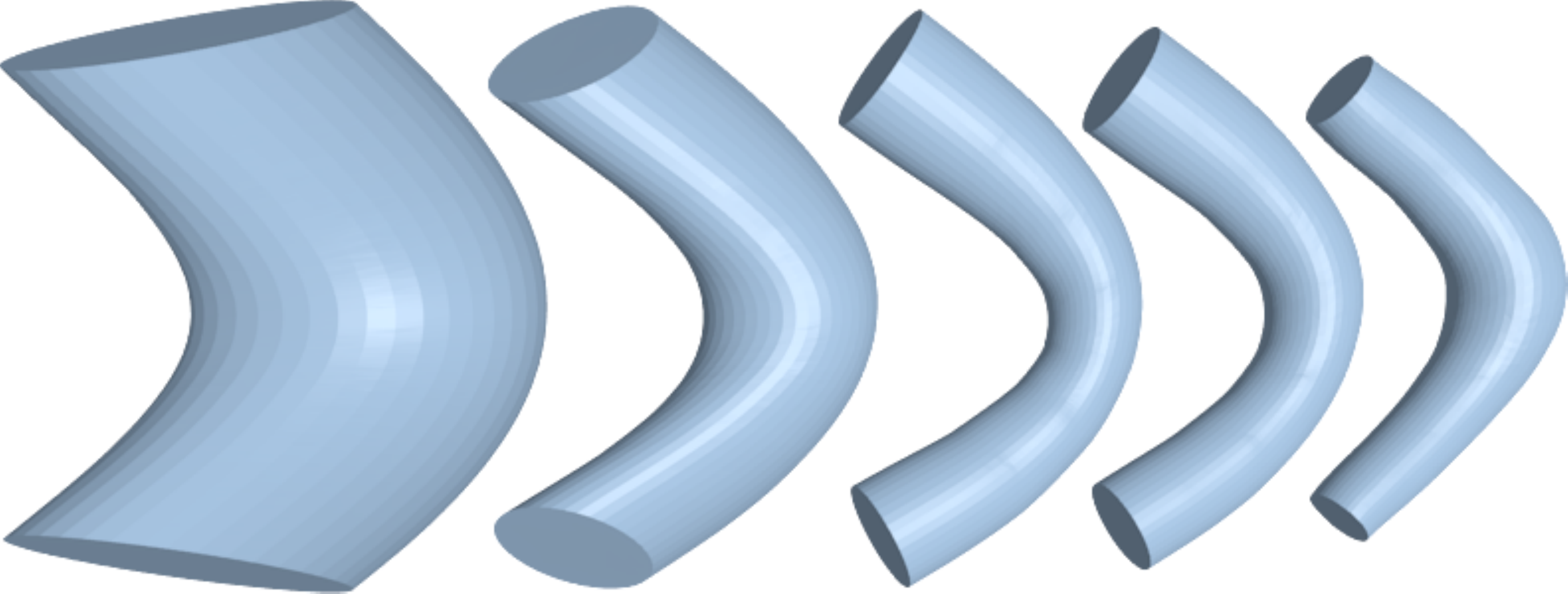}
    \caption{Results subject to different regularization coefficients. Deformed solid cylinder upon three point constraints. From left to right: $\alpha= 0.01, 0.05, 0.1, 1, 10$.}
    \label{fig:penalty} 
\end{figure}

\begin{figure}[t]
    \includegraphics[width=0.45\textwidth]{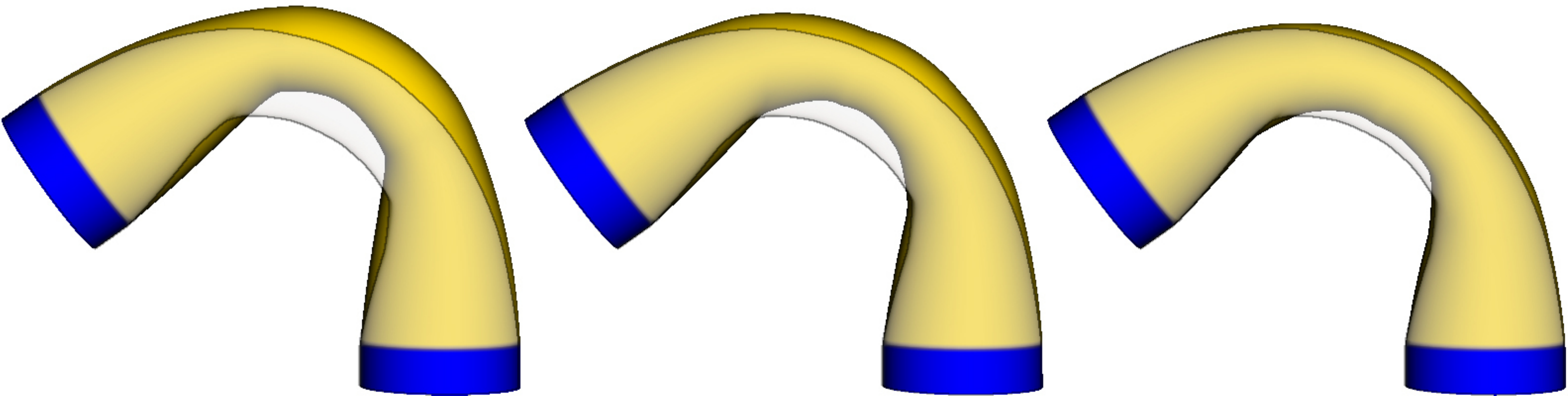}
    \caption{Results with varied number of rotational proxies. From Left to right: $d = 5, 9, 17$: for each, deformed result with 32 rotational proxies is shown with dark contours.}
    \label{fig:moreproxies}
\end{figure}

We divide $r$ local rotations into $d$ rotational clusters spatially,
which is an over-segmentation of input meshes. Rotations within each patch segment
are desired to be similar in deformations. Empirically, we found that a simple k-means clustering 
on weighted Laplacian-Beltrami eigen-vectors fits well with our scheme,
which cuts surface/solid mesh into $d$ patches. 
We revise the original energy by assuming 
\begin{equation}\mathbf r_k \approx \mathbf s_{i_k} + \mathbf q_k\;,
\label{eq:rolinear}\end{equation}
where $\mathbf s_{i_k}\in SO(3)$ denotes $i_k$-th patch-wise frame rotation of the cluster that
the vertex $k$ belongs, and 
\begin{equation}
  \mathbf q_k = \begin{pmatrix}
    q_0 & -q_3 & q_2 \\
  q_3 & q_0 & -q_1 \\
  -q_2 & q_1 & q_0
\end{pmatrix}^{(k)} =  \sum_{i=0}^3 q^{(k)}_i \mathbf d_i\;.
\end{equation}
It should be noted that Laplacian surface editing~\cite{sorkine2004laplacian}
utilizes $\mathbf q_k$ to approximate the rotation matrix, whereas we use $\mathbf q_k$
to approximate the difference of two rotation matrices $\mathbf r_k -\mathbf s_{i_k}$.
It leads to a different energy by appending L2 penalties subject to the regulators $\mathbf q_k$:
\begin{equation}\mathcal E'(\mathbf V', \mathbf Q, \mathbf S) = \mathcal E(\mathbf V', \mathbf S + \mathbf Q) + \alpha \sum_{k=1}^ra_k\norm{\mathbf q_k}^2 +\beta \sum_{k=1}^r a_k\norm{\mathbf q_k \mathbf n_k}^2,
\end{equation}
where $a_k$ denotes the element-wise area/volume,
$\alpha$ denotes the penalty coefficient of overall spatial distortion,
and $\beta$ denotes the additional penalty coefficient of surface normal distortion (if applicable).
$\alpha$ and $\beta$ are empirically chosen.
(See Fig.~\ref{fig:penalty} for deformations subject to different
penalty coefficients $\alpha + \beta$). 
One potential issue is that using this penalty may incur
surface folds when shapes are bended at large angles.
To counteract such effects, we optionally use an extra regularization
term appended to $\mathcal E'(\mathbf V',\mathbf Q, \mathbf S)$ penalizing
the moving frame differentials~\cite{lipman2007volume}, i.e.,
\begin{equation}\label{eq:addreg}
  \gamma \sum_{(k,j)\in \mathcal H}
  a_{k,j}\norm{\mathbf s_{i_k}\!\!+\!\mathbf q_k\!\!-\!\mathbf s_{i_j}\!\!-\!\mathbf q_j}^2,
\end{equation}
where $\mathcal H$ is the set of neighboring local frames and $a_{k,j} = (a_k + a_j)/2$.
The two bending cylinder examples in Fig. \ref{fig:testsuites} are produced by penalizing
the moving frame differentials. 

Let $S_{9d\times 1}$ and $Q_{4r\times 1}$ be the
vectorization of $(\mathbf s_i)$ and $(\mathbf q_k)$ respectively.
$\mathcal E'$ is quadratic in terms of $\mathbf V'$ and $\mathbf Q$,
and its partial gradient w.r.t. $\mathbf V'$ and $\mathbf Q$ is again linear 
in terms of $\mathbf S$. Hence again we can write $\mathcal E'$ as 
\begin{equation}
  E'(V',Q, S) = \frac 12 [V;Q]^T L [V;Q] - S^T M [V;Q]\label{approxenergy}
 + S^TNS + \mbox{constant}\nonumber\;,
\end{equation}
where $S$ are the rotational proxies of our model, $N\neq \mathbf 0$ 
iff the extra regularization \eqref{eq:addreg} is present.

There is an interesting discussion about the difference between 
$\mathcal E$ and $\mathcal E'$, because $\mathcal E'$
includes near-isotropic scaling which has arguable values over distortion
in only one direction for artistic modeling purpose 
in case the desired deformation is far from the isometry
\cite{sorkine2004laplacian,lipman2008green}.
(See Fig.~\ref{fig:arapbad} for comparison with the ARAP energy.)\\

\noindent\textbf{Linear proxies.} Besides rotational proxies,
we add $3m$ linear proxies via pseudo-spatial linear constraints,
\begin{equation}W_{3m\times 3n} V' = X\;,
\end{equation}
where $X$ are the linear proxies of our model.

Intuitively, $W = \widehat W_{m\times n}\otimes I_{3\times 3}$ spans a finite dimensional 
linear space to approach the uncertainty set of onsite constraints provided by users.
Its choice reflects how we reduce the dimension of anticipated constraints, 
as suggested by the use of variational subspace, 
A simple choice is a sparse sampling of $m$ vertices
(shown as Fig.~\ref{fig:facemodeling}), i.e (under a permutation)
\[ W_{3n\times 3m} = \{\ldots;\overbrace{0,\ldots,0,1,0,\ldots,0}^{\mbox{\small sample at singe vertex i}};\ldots\}_{n\times m}\otimes I_{3\times 3},\]
and an alternative one is to utilize $m$ vertices groups via clustering,
i.e., (under a permutation)
\[ W_{3n\times 3m} = \dfrac{1}{\mathcal{N}}\{\ldots;\underbrace{\ldots,0,\!\!\!\!\!\!\!\!\!\!\!\!\!\overbrace{1,\ldots,1}^{\mbox{\small vertices group j of size $\mathcal{N}$}}\!\!\!\!\!\!\!\!\!\!\!\!\!,0,\ldots}_{\mbox{\small j-th rows}};\ldots\}_{n\times m}\otimes I_{3\times 3}.\]

{To this point, technically contrast with our approach, 
standard model reduction technique employ a strategy that vertices are explicitly represented 
in low-order by $V'_{3n\times 1} = K_{3n\times 3m} X_{3m\times 1}$. 
In order to compute a reasonable subspace, 
different smoothness criterion are exposed on computing $K$, 
such as heat equilibrium\cite{baran2007automatic},
exponential propagating weights\cite{jacobson2012fast}, 
biharmonic smoothness\cite{jacobson2011bounded}.} 
We instead reduce the dimension of constraints, and
the subspace are then automatically solved accordingly. \\

\noindent\textbf{Variational subspace. } With context of approximated energy $E'$,
we are to solve the linear variational problem so as to derive a reduced representation of $V'$
in terms of proxies $S$ and $X$, i.e.,
\begin{equation}\begin{array}{rl}
\min\limits_{V',Q} & E'(V', Q, S)\\
\mbox{s.t.} & W_{3m\times 3n} V' = X\;.
\end{array}
\end{equation}

By KKT condition introducing Lagrange multipliers $\Lambda$,
we have a set of linear equations in respect of $ V',  Q, \Lambda$,
which can be expressed as matrix form 
\begin{equation}\label{eq:KKTsubspace}
  \begin{pmatrix} L &  W^T\\
  W &  0
\end{pmatrix}
[V'; Q; \Lambda] =  [M^T  S; X]\;.
\end{equation}
This then implicitly establishes a linear map
\begin{equation}\label{eq:ARAPsubspace}
  \begin{array}{rcl}
  [V'; Q]
  & = &N_{W} X + U_{W} S\;,
\end{array}
\end{equation}
where each column of matrices $N_{W}, U_{W}$ can be pre-computed
by a sparse linear solver with a single preconditioning (LU or Cholesky), subject to each single
variable in vector $X$ and $S$. 
Solving for $N_W$ and $U_W$ only need one time computation in the offline stage.\\

\noindent\textbf{Sub-manifold integration.}
Provided variational subspace, $X$ and $S$ span a sub-manifold of deformations. We then restrict our scope to determine reduced variables $X$ and $S$.  We employ a routine similar to alternating least square~\cite{sorkine2007rigid}, where we alternatively update $X$ and $S$ via two phases. \\

\noindent\textbf{Phase 1: provided $S^{(i)}$, solve for $X^{(i)}$}.

By substituting \eqref{eq:ARAPsubspace} into approximated ARAP
energy \eqref{approxenergy}, we derive a reduced ARAP energy as 
\begin{equation}\label{eq:ARAPreduced}
  E''(S, X) = \frac 12  X^T \widetilde{L} X -  S^T \widetilde{M} X + \mbox{constant}\;.
\end{equation}
where $\widetilde{L} = ( N_{W})^T L  N_{ W}$
and $\widetilde{ M} = M  N_{W} + (U_{W})^T L  N_{W}$.
With onset hard constraints $W_{\mbox{\scriptsize eq}} V' = P_{\mbox{\scriptsize eq}}$ specified by the user
(where $W_{\mbox{\scriptsize eq}}$ are positional constraints and  
$P_{\mbox{\scriptsize eq}}$ are their values),
we are then to solve for linear proxies $X$ 
\begin{equation}\label{eq:reduced_hard}
  \begin{array}{rl}
    \min\limits_{X} &E''(S^{(i)},X) \\
    \mbox{s.t.} &  N_{\mbox{\scriptsize eq}}   X  = P_{\mbox{\scriptsize eq}} - U_{\mathrm{eq}}  S^{(i)}    
\end{array}
\end{equation}
 where $ N_{\mbox{\scriptsize eq}}=W_{\mbox{\scriptsize eq}} N_{W}, U_{\mbox{\scriptsize eq}} = W_{\mbox{\scriptsize eq}} U_{W}$. Hard constraints are the default setting of our framework.

Alternatively, we can pose on-site soft constraints as
\begin{equation}\label{eq:reduced_soft}
  \begin{array}{rl}
    \min\limits_{X} &E''(S^{(i)},X) + \delta \norm{N_{\mbox{\scriptsize eq}}   X  + U_{\mathrm{eq}}  S^{(i)} - P_{\mbox{\scriptsize eq}}}^2,    
\end{array}
\end{equation}
where $\delta\!\!>\!\!0$ is adjusted interactively by user to match the desired effects.
We input $W_{\mbox{\scriptsize eq}}, P_{\mbox{\scriptsize eq}}$ and 
solve the integrated reduced model interactively.

Remark that because optimization problems (equations \eqref{eq:reduced_hard}
and \eqref{eq:reduced_soft}) are again linear variational,
it can be efficiently solved by a standard dense linear solver:
(1) pre-computing LU factorization of matrix (not related to $S$)
at the stage to specify constraint handlers $W_{\mbox{\scriptsize eq}}$, and
(2) backward substitution on the fly at the stage to drag/rotate handler.\\

\noindent\textbf{Phase 2: provided $X^{(i)}$ and $S^{(i)}$, compute $S^{(i+1)}$}.
Rather than minimizing the reduced energy functional $E''$ (shown in Eq. \eqref{eq:ARAPreduced})
in terms of $S$, we instead want rotational clusters to adapt for the existing deformation.
Letting $[V'^{(i)} ; Q^{(i)}] = N_W X^{(i)} + U_W S^{(i)}$, 
we fit a patch-wise local frame of rotational clusters subject to deformed mesh $V'^{(i)}$
by dumping relations $Q^{(i)}$ and their penalties
$\alpha\!\!=\!\! \beta\!\! =\!\! \gamma\!\! =\!\! 0$,
and optimizing a simplified energy
$\mathcal E(\mathbf V'^{(i)}, \mathbf S) = E'(V'^{(i)},\mathbf 0,S)$, which is equivalent to 
\begin{equation}
  \begin{array}{rl}
    \max\limits_{S} & S^T M [V'^{(i)};\mathbf 0] = S^T (M_{N} X^{(i)}+ M_U S^{(i)})\\
    \mbox{s.t.} & \mathbf s_i \in SO(3), \quad i=1,\ldots,d,
  \end{array}
\end{equation}
where $M_N$ and $M_U$ are pre-computed. It is well known that
those rotation fittings can be solved in parallel via singular value decomposition of
each gradient block of $\mathbf s_i$. For $3\times 3$ matrix,
we employ the optimized SVD routines by McAdams and colleagues~\cite{mcadams2011efficient}
that avoid reflection, i.e., guarantee the orientation.\\


\subsection{Algorithm overview}
\begin{figure*}[ht]
  \includegraphics[width=\textwidth]{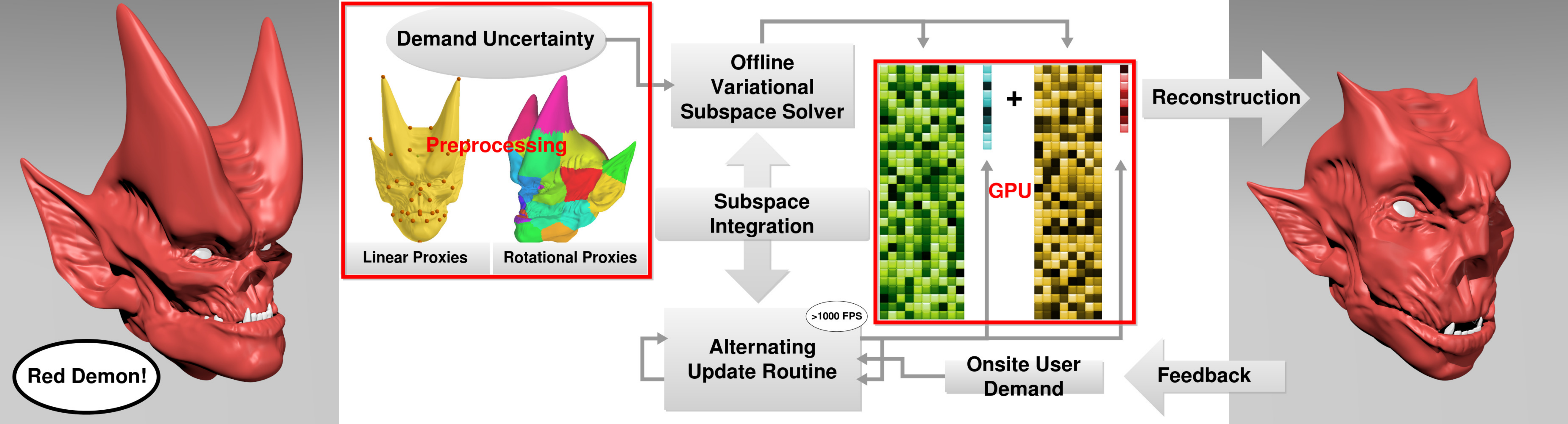}
  \caption{Interactive mesh modeling framework of our approach. From left to right:
    (1) Input red demon model $V$;
    (2) Pre-process to set pseudo-constraint points as linear proxies
    and near-rigid parts as rotational proxies;
    (3) Compute variational subspace $N_W, U_W$ offline, and load into display device.
    (4) Prepare subspace integration $\widetilde{L}$, $\widetilde{M}$, $M_N$ and $M_U$.
    (5) At run-time, given on-site user demand $W_{\mbox{\scriptsize eq}}, P_{\mbox{\scriptsize eq}}$,
    solve for reduced variables $X, S$
    and upload to display device (bandwidth saving);
    (6) Display deformed mesh and feedback.
  }
  \label{fig:facemodeling}
\end{figure*}

We review our previous mathematical formulations, and summarize our algorithm
into three stages (see also Fig.~\ref{fig:facemodeling}):

\noindent\textbf{Pre-compute.} The user loads initial mesh model $\mathcal M$,
linear proxies $W$, rotational proxies $(i_k)$, and affine controllers (if applicable).
Our algorithm constructs a sparse linear system to solve for variational subspace
$N_W$ and $U_W$, and then pre-computes $\widetilde{L}$, $\widetilde{M}$, $M_N$ and $M_U$.\\

\noindent\textbf{Prepare on-site constraints.} When above pre-computed matrices are present,
the user can only freely specify the intended constraint handler on-site.
They are in the form of $W_{\mbox{\scriptsize eq}}$. Our algorithm then proceeds to
compute $N_{\mbox{\scriptsize eq}}$ and $U_{\mbox{\scriptsize eq}}$, and pre-factorize the
linear system (see equation \eqref{eq:reduced_hard} or \eqref{eq:reduced_soft}).
If a user introduces a brand new set of constraints on-site, this stage will be re-computed within tens of milliseconds.
The timing regard to different settings has been reported in 
Table~\ref{table:timing} column ``OP''.  \\

\noindent\textbf{Deform on the fly.} 
Our algorithm allows the user to deform meshes
on the fly, which means the user can view the deformation results instantly
by controlling constraint handlers. 
For each frame, our model takes in positional constraints $P_{\mbox{\scriptsize eq}}$, 
calls an alternating routine (with global rotation adaption) interactively to solve for
proxy variables $X$ and $S$, and reconstructs and displays the deformed mesh.
To guarantee real-time performance, we used a fixed number of iterations per frame.
By initializing an alternating routine with the previous frame proxies, we do not
observe any disturbing artifacts even when using only 8 iterations.\\

\subsection{Results and Discussion}\label{sec:results}
\begin{table*}[ht]
  \centering{\small
  \begin{tabular}{|c|c|c|c|c|c|c|c|c|c|c|c|c|}\hline
    & \multicolumn{2}{c|}{Input Model} & \multicolumn{2}{c|}{Proxies}& \multicolumn{3}{c|}{Runtime} &\multicolumn{2}{c|}{Pre-computation} &\\\hline
    Model & Vert. & Type & Linear & Rot. & \parbox{1cm}{1 Iter.\\ ($\mu$s)} & 
    \parbox{.8 cm}{Df.\\ (ms)} & \parbox{.8 cm}{Total\\ (ms)} & \parbox{1cm}{Subspace\\ (s/GB)} & 
    \parbox{.8 cm}{OP \\ (ms)} & Fig.\\\hline
    Cylinder & 5k & Tri. & 33 & 12 & 50 & 1.4 & 2 & 14 (0.3) & 14 & \ref{fig:testsuites}\\\hline
    Cactus & 5k & Tri. & 33 & 27 & 90 & 2 & 3 & 18 (0.4) & 16 & \ref{fig:testsuites} \\\hline
    Bar & 6k & Tri. & 33 & 52 & 93 & 3.8 & 4.8 & 36 (.5) & 20 & \ref{fig:testsuites} \\\hline
    Bumpy Plane & 40k & Tri. & 33 & 27 & 85 & 14 & 15 & 200 (2.7) & 38 & \ref{fig:testsuites} \\\hline
    Plate Box & 4k & Tet. & 25 & 25 & 62 & 2 & 2.7 & 30(.4) & 11 & \ref{fig:surfacevssolid} \\\hline
    Solid Cylinder & 8k & Tet. & 33 & 52 & 110 & 4& 5 & 90 (.8) & 22 & \ref{fig:penalty} \\\hline
    Tree & 3.6k & Tri. & 60 & 60 & 137 & 3.5 & 5 & 34(.4) & 10 & \ref{fig:header}\\\hline
    Fertility & 25k & Tet. & 29 & 26 & 78 & 5.2 & 6 & 148 (2.4) & 28 & \ref{fig:header}\\\hline
    Dinosaur & 21k & Tri. & 46 & 34 & 108 & 12 & 14 & 115 (1.7)& 24 & \ref{fig:dino}\\\hline
    Dragon & 53k & Tri. & 20 & 20 & 55 & 14 & 15 &  198 (2.7)& 64 & \ref{fig:conformal}\\\hline
    Red Demon &  80k & Tri. & 30 & 28 & 90 & 35 & 36 & 498 (5.8) & 106 & \ref{fig:facemodeling}\\\hline
  \end{tabular}
  \caption{Model statistics and serial performance
    on a HP laptop with an Intel i7 2.20GHz $_{\times 8}$ Processor.
    From left to right:
    number of vertices,
    type of mesh,
    number of linear proxies,
    number of rotational proxies,
    time in $\mu$ seconds for one iteration,
    time in milliseconds for mapping reduced solution info full space
    (Intel MKL on CPU performance),
    time in milliseconds for full optimization per frame, 
    time in seconds (and memory in GBytes) for pre-computation of subspace,
    time in milliseconds for computation of on-site preconditioner,
    figure that shows the configuration.}
  \label{table:timing}
}
\end{table*}

\begin{figure}[t]
    \includegraphics[width=0.45\textwidth]{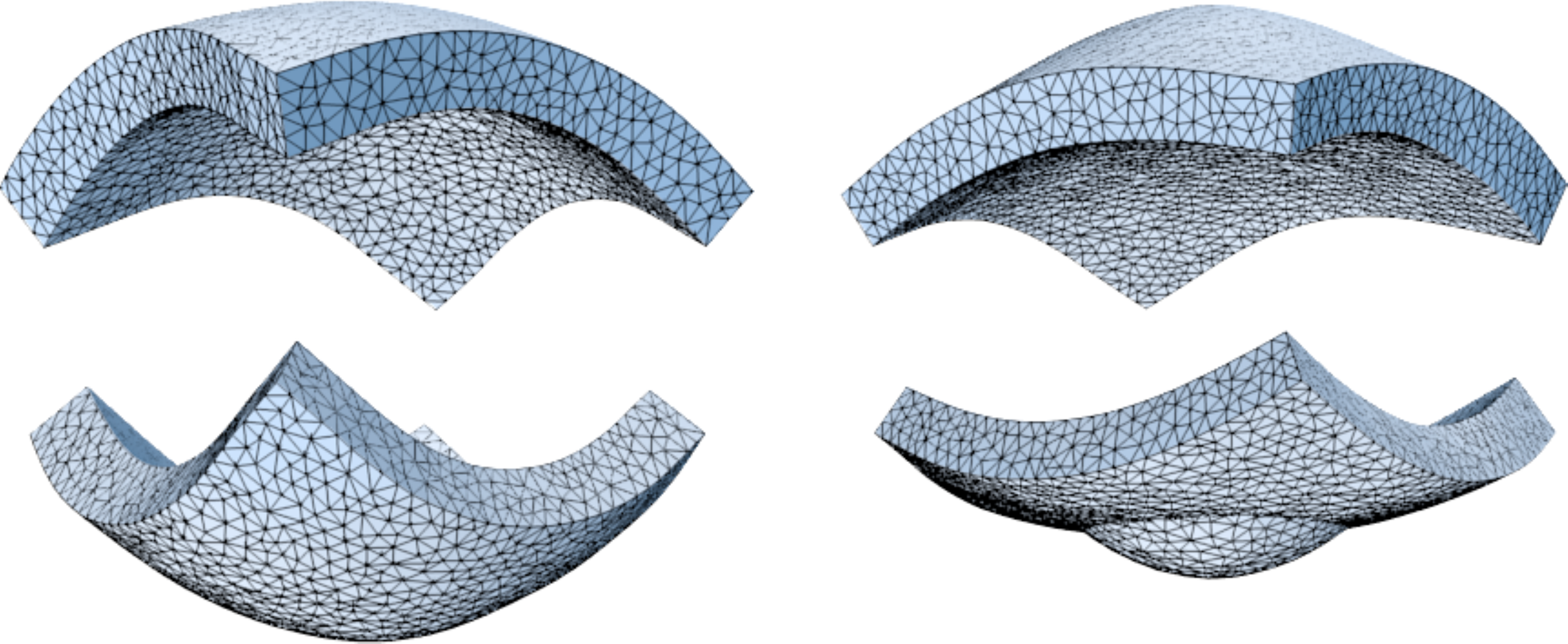}
    \caption{Our reduced model preserves the nature of different functions. Left: volumetric deformed mesh; Right: deformed surface (self-occlusion possible) under same constraints.}
    \label{fig:surfacevssolid}
\end{figure}
\begin{figure}[t]
    \includegraphics[width=0.45\textwidth]{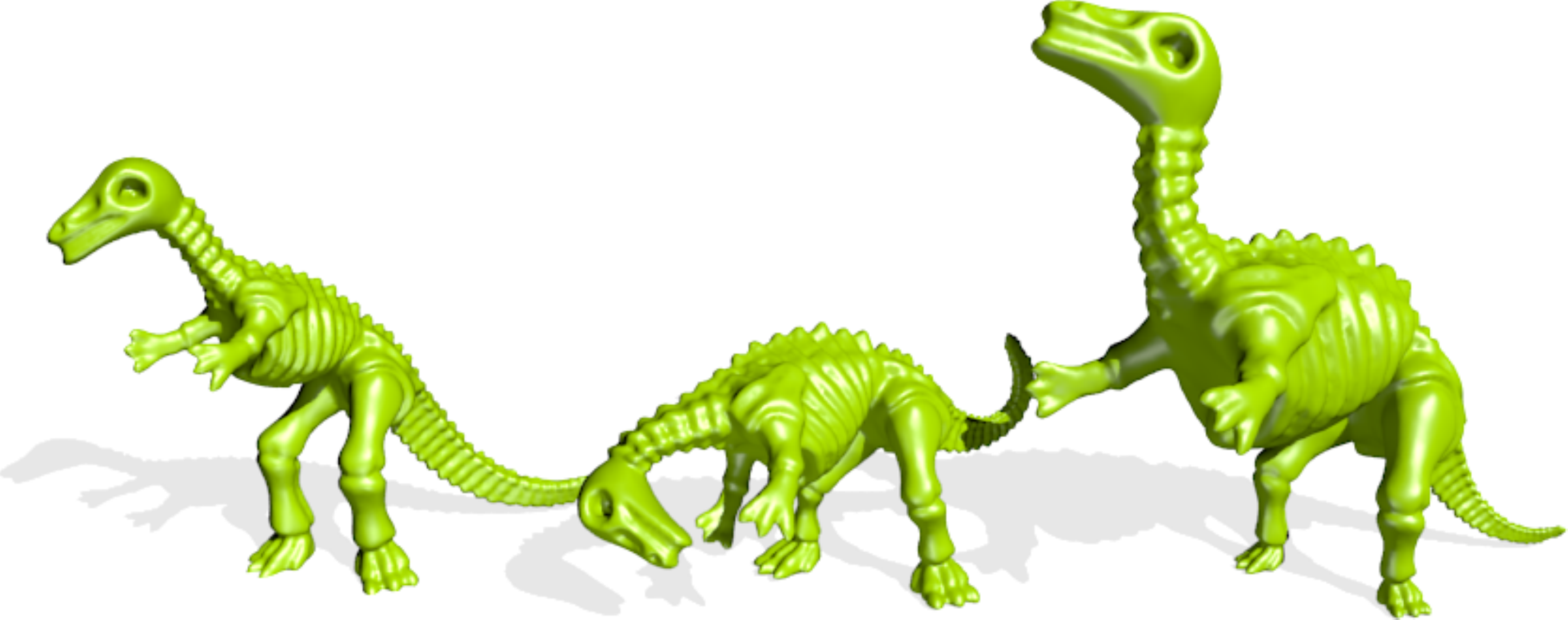}
    \caption{{Our revised energy does not reveal linearization artifacts when the deformation is isometric (middle). It also favors isotropic scaling when the model is stretched or shrunk (right). The original model is shown in the left.}}
    \label{fig:dino}
\end{figure}

\begin{figure*}[ht]
\centering{
  \includegraphics[width=.95\textwidth]{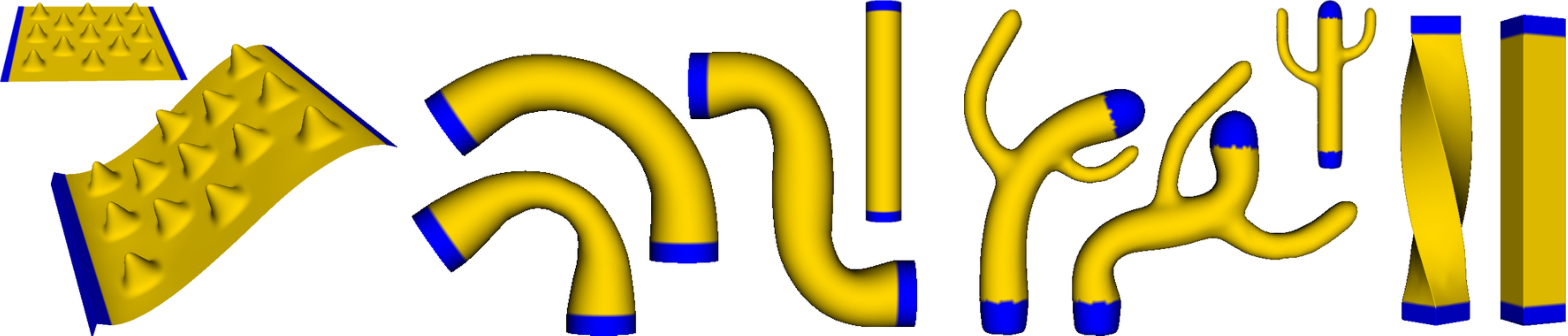}
  \caption{Approximation quality of our method in all images
    is demonstrated on the test suite of models
    introduced in [Botsch and Sorkine 2008];
    From left to right: Bumpy Plane, Cylinder, Cactus, Bar.
    Multiple types of deformations, including bending, shifting and twisting,
    are tested.}
  \label{fig:testsuites}
  }
\end{figure*}

We implement our framework for deformable mesh modeling and demonstrate our
results by examples which include standard deformation suites
introduced in~\cite{botsch2008linear}.
Results of our approach on a set of typical test meshes are shown
in Fig.~\ref{fig:testsuites}).
The results shown can be compared with results of high quality methods without
model reduction, including PriMo\cite{botsch2006primo}, also shown in~\cite{botsch2008linear}.
Besides, we also demonstrate the strength of our method in conformal setting,
where we configure scaling factors in our modeling framework
(see Fig.~\ref{fig:conformal}). In our experiments, the modeling framework
runs robustly on various models, for different types of transformation, such as
small and large rotations, twisting, bending, and more 
as shown in Fig.~\ref{fig:cactus}. It works reasonably naturally on
both surface and solid meshes, in which user's choice of energy
controls the desired behaviors (see Fig.~\ref{fig:surfacevssolid}).
It also accommodates different hyper-parameter setting, such as the number and type of proxies,
to produce predictable and reasonable results (see Fig.~\ref{fig:moreproxies}).

Based on our CPU implementation, We report timing of our algorithm working
on different models presented in our paper in Table~\ref{table:timing}.
All timing results are generated on an Intel Core$^{\mbox{\scriptsize TM}}$ i7-2670QM 2.2GHz $\times$8
processor with 12 GB RAM.
It has been shown that the time used in reduced model iteration
is not related to the geometric complexity, and the overall computation per frame
is magnitude faster than that as reported in \cite{hildebrandt2011interactive}.
The computational framework used in our paper 
is almost as same as the one used in~\cite{jacobson2012fast},
thus the performances are comparable.
It is also shown that the process of mesh reconstruction, which is a matrix-vector product
(see Eq.~\eqref{eq:ARAPsubspace} and column ``Df.'' of Table~\ref{table:timing}),
is the bottle-neck of overall computation, yet it is embarrassingly parallel.

Comparing to other cell-based model reduction
methods~\cite{sumner2007embedded,botsch2007adaptive}, our approach utilizes
a much smaller number of reduced variables. Typically, we adopt no more than
35 linear proxies, and no more than 60 rotational proxies. 
Besides, the configuration of proxies gives user the freedom to design
his/her own needs in modeling a particular mesh. Instead restricting variability in modes
and modal derivatives space~\cite{hildebrandt2011interactive}, artist, based on his intentions,
can cut shape into near rigid parts (each for a rotational proxy),
and specify pseudo constrain locations as linear proxies. Fig.~\ref{fig:facemodeling}
demonstrates a modeling scenario where artist intended to adjust
mouth, nose and eyes on a face model: semantic parts are in first annotated,
variational reduced deformable model is then pre-computed for on-line editing.\\

\section{Theory of Variational Subspace}
The notations used follows the preliminary setup in section~\ref{sec:setup}.

\subsection{Concept of Approximation}

\begin{definition}[Variational Subspace]
  Given a quadratic programming problem in the form of
  Eq.~{\eqref{eq:quadprog}}, choose $C$ and $D$ for the problem in the
  two-stage form as Eq.~{\eqref{eq:2stage2}}. The solution $X^{\ast}  (Z,Y;C,D)$ 
  for the first stage problem is hence given by Eq.~{\eqref{eq:qpsubspace}}. 
  The subspace spanned by columns of $N$ and $U \cdot D$ are called variational subspace.
\end{definition}

\begin{proposition}
  $N$ is a matrix of size $n \times d$ and $U \cdot D$ is a matrix of size $n \times k$. 
  Their columns span a linear subspace $\tmop{Span}(U) + \tmop{Span}(U\cdot D)$
  where the reduced solution belongs. 
  Those columns are computed by solving a variational formulation 
  provided by Eq.~{\eqref{eq:qpform}}.
\end{proposition}

In Eq.~\ref{eq:quadprog}, $H$ is only guaranteed to be positive
semi-definite. We have Cholesky decomposition $H=L^{T} L$, where $L$ is
upper triangular with non-negative diagonal entries~{\cite{golub2012matrix}}.
Denote the pseudo-inverse of $L$ by $L^{+}$, then for any $X \in \mathbb{R}^{n}$, 
we have a two-part orthogonal decomposition $X= \tilde{X}+ \bar{X}$, where
\begin{equation}
  \label{eq:2partdecom} \tilde{X} =L^{+} LX, \hspace{1em} \bar{X} = (I-L^{+}
  L) X.
\end{equation}
\begin{proposition}
  With the two-part decomposition Eq.~{\eqref{eq:2partdecom}}, we have
  \begin{equation}
    L \tilde{X} =LX, \hspace{1em} L \bar{X} =0, \hspace{1em} \tilde{X}^{T} 
    \bar{X} =0.
  \end{equation}
  In addition, if $H=L^{T} L$ is positive definite, $\bar{X} =0$.\\
\end{proposition}

Moreover, we can rewrite the optimization problem Eq.~{\eqref{eq:quadprog}} as
\begin{equation}
  \begin{array}{rl}
    \min_{X} & (1/2)  ( \tilde{X} + \bar{X} )^{T} L^{T} L ( \tilde{X} +
    \bar{X} ) -q^{T}  ( \tilde{X} + \bar{X} )\\
    \text{s.t.} & A^{T}  ( \tilde{X} + \bar{X} ) =b\\
    & \tilde{X} =L^{+} LX, \hspace{1em} \bar{X} = (I-L^{+} L) X.
  \end{array}
\end{equation}
Since $L \bar{X} =0$, we simplify above formulation to
\begin{equation}
  \label{eq:originqp} \begin{array}{rl}
    \min_{X} & f (X) = (1/2)  \tilde{X}^{T} L^{T} L \tilde{X} -q^{T} 
    \tilde{X} -q^{T}  \bar{X}\\
    \text{s.t.} & A^{T}  \tilde{X} =b-A^{T}  \bar{X}\\
    & \tilde{X} =L^{+} LX, \hspace{1em} \bar{X} = (I-L^{+} L) X.
  \end{array}
\end{equation}
It is observed that only $\tilde{X}$ appears in second-order 
term in the objective function of Eq~\eqref{eq:originqp}. Suppose the optimal solution to
Eq.~{\eqref{eq:quadprog}} is $X_{\min}$ with a two-part decomposition (given
by Eq.~{\eqref{eq:2partdecom}}) $X_{\min} = \tilde{X}_{\min} +\bar{X}_{\min}$,
we then consider the following companion optimization problem
\begin{equation}
  \label{eq:eqquadprogbak} \begin{array}{rl}
    \min_{\tilde{X}} & \tilde{f} ( \hat{X} ) = (1/2)  \tilde{X}^{T} L^{T} L
    \tilde{X} -q^{T}  \tilde{X} -q^{T}  \bar{X}_{\min}\\
    \text{s.t.} & A^{T}  \tilde{X} =b-A^{T}  \bar{X}_{\min}\\
    & (I-L^{+} L)  \tilde{X} =0.
  \end{array}
\end{equation}
Remark $-q^{T} \bar{X}_{\min}$ which appears in objective function 
of Eq.~\eqref{eq:eqquadprogbak} is a constant. 
We see Eq.~{\eqref{eq:eqquadprogbak}} can be equivalently solved in two steps: In
the first step, we solve the following problem
\begin{equation}
  \label{eq:eqquadprog} \begin{array}{rl}
    \min_{\tilde{X}} & \tilde{f} ( \tilde{X} ) +q^{T}  \bar{X}_{\min} = (1/2) 
    \tilde{X}^{T} L^{T} L \tilde{X} -q^{T}  \tilde{X}\\
    \text{s.t.} & A^{T}  \tilde{X} =b-A^{T}  \bar{X}_{\min}.
  \end{array}
\end{equation}
And in the second stage, we project the solution to $L^{+} L \tilde{X}_{\min}$,
where $\tilde{X}_{\min}$ is the solution to Eq.~{\eqref{eq:eqquadprog}}.
\begin{theorem}
  Suppose $X_{\min}$ is the unique solution to Eq.~{\eqref{eq:originqp}},
  $\tilde{X}^{\circ}_{\min}$ is the unique solution to
  Eq.~{\eqref{eq:eqquadprogbak}}, and $\tilde{X}_{\min}$ is the unique solution to
  Eq.~{\eqref{eq:eqquadprog}}, then we have 
  $\tilde{X}_{\min}^{\circ} =L^{+}LX_{\min} =L^{+} L \tilde{X}_{\min}$.  
  \begin{proof}
    We observe that $L^{+} LX_{\min}$ satisfies the
    constraint of Eq.~{\eqref{eq:eqquadprogbak}}, and objective
    functions of Eq.~\eqref{eq:originqp} and Eq.~\eqref{eq:eqquadprogbak} coincide, 
    i.e., $f (X_{\min} ) = \tilde{f} (L^{+} LX_{\min} )$.
    Therefore, as $\tilde{X}^{\circ}_{\min}$ is the minimizer to
    Eq.~{\eqref{eq:eqquadprogbak}}, 
    we have \[f (X_{\min} ) \ge \tilde{f} ( \tilde{X}^{\circ}_{\min}).\] 
    On the other hand, $\tilde{f} ( \tilde{X}^{\circ}_{\min}) =f ( \tilde{X}^{\circ}_{\min} + \bar{X}_{\min} )$. 
    Given $X_{\min}$ is the minimizer to Eq.~{\eqref{eq:originqp}},
    we also have \[\tilde{f} (\tilde{X}^{\circ}_{\min} ) \ge f (X_{\min} ).\] 
    Hence, the equality holds for $\tilde{f} ( \tilde{X}^{\circ}_{\min} ) =f (X_{\min} )$. 
    Given that the optimum exists and is unique, we have 
    \[\tilde{X}^{\circ}_{\min} +\bar{X}_{\min} =X_{\min} \Rightarrow \tilde{X}^{\circ}_{\min} =L^{+} LX_{\min}.\]
    
    The proof of $\tilde{X}_{\min}^{\circ} =L^{+} L \tilde{X}_{\min}$ is similar: observe that
    $\tilde{f} ( \tilde{X}^{\circ}_{\min} ) = \tilde{f} (L^{+} L \tilde{X}_{\min})$ 
    and $L^{+} L \tilde{X}_{\min}$ satisfies the constraint of
    Eq.~{\eqref{eq:eqquadprogbak}}.
  \end{proof}
\end{theorem}

\begin{definition}
  Two solutions subject to the form Eq.~{\eqref{eq:quadprog}} 
  (parameterized by $A$, $q$, and $b$) are called \textit{quotient equivalent}, 
  if they share the same companion problem defined
  by Eq.~{\eqref{eq:eqquadprog}}, i.e. their $\hat{b}(A,b,q) =b-A^{T} \bar{X}_{\min}$ 
  are the same. This forms group equivalence in the space of solutions.
\end{definition}

For example, let $H$ be the Laplacian operator $\Delta$, the solution would
minimize the the L2 norm of first-order gradient. In such case, two problems
are of quotient equivalence if their optimal solutions preserve to an additive
constant. We use the distance between two solution groups under the quotient
equivalence to measure the approximation error. 
It is the Mahalanobis distance provided by $H$, i.e.
\[ d_{H} (x,y) = (x-y)^{T} H (x-y) = (Lx-Ly)^{T}  (Lx-Ly) . \]
Let $LX= \hat{X}$, $\hat{q} =L^{+} q$, $\hat{A} =L^{+} A$, and $\hat{b}=b-A^{T}  \bar{X}_{\min}$,
we rewrite Eq.~{\eqref{eq:eqquadprog}} (but not equivalent) as
\begin{equation}
  \label{eq:eq2quadprogbak} \begin{array}{rl}
    \min_{\hat{X}} & (1/2)  \hat{X}^{T}  \hat{X} - \hat{q}^{T}  \hat{X}\\
    \text{s.t.} & \hat{A}^{T}  \hat{X} = \hat{b}\\
    & (I-LL^{+} )  \hat{X} =0.
  \end{array}
\end{equation}
Similar to the treatment of Eq.~{\eqref{eq:eqquadprogbak}}, we can in first solve
\begin{equation}
  \label{eq:eq2quadprog} \begin{array}{rl}
    \min_{\hat{X}} & (1/2)  \hat{X}^{T}  \hat{X} - \hat{q}^{T}  \hat{X}\\
    \text{s.t.} & \hat{A}^{T}  \hat{X} = \hat{b} .
  \end{array}
\end{equation}
and then project the solution to $LL^{+}  \hat{X}_{\min}$, where $\hat{X}_{\min}$
is the solution to Eq.~{\eqref{eq:eq2quadprog}}.

Since we are always interested in distance measure $d_{H}$ for different
solutions, the projection step in solving Eq.~{\eqref{eq:eqquadprogbak}} is
not necessary to compute $d_H$. The distance between two solution groups
$X_{1}$ and $X_{2}$ of the quotient equivalence is therefore the Euclidean distance
between $\hat{X}_{1} =LX_{1}$ and $\hat{X}_{2} =LX_{2}$.\\

Similar to the treatment of Eq.~{\eqref{eq:quadprog}} and
Eq.~{\eqref{eq:eq2quadprog}}, we can derive the two-stage problem from
Eq.~{\eqref{eq:2stage}} as
\begin{equation}
  \label{eq:eq2stage} \begin{array}{rl}
    \min_{\hat{X}} & (1/2)  \hat{X}^{T}  \hat{X} - \hat{Y}^{T}  \hat{D}^{T} 
    \hat{X}\\
    \text{s.t.} & \hat{C}^{T}  \hat{X} = \hat{Z}\\
    & A_{C}^{T}  \hat{Z} = \hat{b} ,
  \end{array}
\end{equation}
where $\hat{D} = (L^{+} )^{T} D$ and $\hat{C} = (L^{+} )^{T} C$. The KKT
condition of its first-stage problem is given similarly as
\begin{equation}
  \left(\begin{array}{cc}
    I & \hat{C}\\
    \hat{C}^{T} & 0
  \end{array}\right) \left(\begin{array}{c}
    \hat{X}^{\ast}\\
    \Lambda
  \end{array}\right) = \left(\begin{array}{c}
    \hat{D}  \hat{Y}\\
    \hat{Z}
  \end{array}\right) . \label{eq:eqqpform}
\end{equation}
where $\Lambda$ is a Lagrange multiplier. We have the following justifications to
only study Eq.~{\eqref{eq:eq2quadprog}} and Eq.~{\eqref{eq:eq2stage}}.\\

\begin{proposition}
  \label{equivalence} If $X_{\min}$ is the optimal solution to
  Eq.~\eqref{eq:quadprog} and $\hat{X}_{\min}$ the optimal solution to
  Eq.~\eqref{eq:eq2quadprog} with $\bar{X}_{\min} = (I-L^{+} L) X_{\min}$ and
  $\hat{b} =b-A^{T}  \bar{X}_{\min}$, we have $LX_{\min} =LL^{+}  \hat{X}_{\min}$. 
  The similar argument also holds for Eq.~{\eqref{eq:2stage}} and Eq.~{\eqref{eq:eq2stage}}.
  
  \begin{proof}
    Because $L^{+} LX_{\min}$ is the optimal solution to
    Eq.~{\eqref{eq:originqp}}, we have $L^{+} LX_{\min} =L^{+} L\tilde{X}_{\min}$, 
    where $\tilde{X}_{\min}$, as mentioned, is the optimal
    solution to Eq.~\eqref{eq:eqquadprog}. On the other hand, we know 
    $L\tilde{X}_{\min} =LL^{+}  \hat{X}_{\min}$. 
    Therefore, we have $L (L^{+} LX_{\min} -L^{+}  \hat{X}_{\min} ) =0 \Rightarrow LX_{\min} =LL^{+} \hat{X}_{\min}$.
  \end{proof}
\end{proposition}

\begin{proposition}
  Suppose $X^{\ast} =NZ+UDY$ is the solution of Eq.~{\eqref{eq:qpform}} and
  $\hat{X}^{\ast} = \hat{N}  \hat{Z} + \hat{U}  \hat{D}  \hat{Y}$ is the
  solution of Eq.~{\eqref{eq:eqqpform}}, we have $LX^{\ast} = \hat{X}^{\ast}$
  if $\hat{Z} =Z-C^{T}  (I-L^{+} L) X^{\ast}$ and $\hat{Y} =Y$.
  
  \begin{proof}
    We have $(L^{+} )^{T}  (L^{T} LX^{\ast} +C \Lambda -DY) =0 \Rightarrow LX^{\ast} + \hat{C} \Lambda - \hat{D} Y=0$ 
    and $C^{T} L^{+} LX^{\ast} =Z-C^{T}  (I-L^{+} L) X^{\ast} \Rightarrow \hat{C}^{T} LX^{\ast} = \hat{Z}$. 
    Therefore, $LX^{\ast}$ satisfies Eq.~{\eqref{eq:eqqpform}}.
  \end{proof}
\end{proposition}

\begin{definition}[Variational Subspace Under Quotient Equivalence]
  Given $\hat{X}^{\ast} = \hat{N}  \hat{Z} + \hat{U}  \hat{D}  \hat{Y}$ to be
  the solution to Eq.~{\eqref{eq:eqqpform}}, the columns of $\hat{N}$ and
  $\hat{U} \cdot \hat{D}$ span the variational subspace 
  $\tmop{Span}(\hat N) + \tmop{Span}(\hat U \cdot \hat D)$ for solutions of
  optimization problem in the form of
  Eq.~{\eqref{eq:eq2quadprog}}.\label{dfvarsubspace}
\end{definition}

The main problem is thus revealed: how close are the exact solution of
Eq.~{\eqref{eq:eq2quadprog}} and subspace solution restricted in a
variational subspace defined by Def.~\ref{dfvarsubspace}, where the closeness
is measured by the Mahalanobis distance $d_{H}$ provided by $H=L^{T} L$.

\subsection{Bound of Approximation Error}\label{sec:bound}
In this section, we provide the proof that the approximation error of model
reduction by variational subspace can be bounded in $d_{H}$.

\begin{proposition}[Exact Solution]
  \label{exactsol} Assuming Eq.~{\eqref{eq:eq2quadprog}} has a unique solution
  which has finite optimum, such that $\hat{A}$ is a full-rank matrix. the
  solution is
  \begin{equation}
    \label{eq:exactsol} \hat{X}_{\min} = ( I- \hat{A}  \hat{A}^{+} )  \hat{q}
    + ( \hat{A}^{+} )^{T} \hat{b} ,
  \end{equation}
  where $\hat{A}^{+} = ( \hat{A}^{T}  \hat{A} )^{-1}  \hat{A}^{T}$ is the
  pseudo-inverse of $\hat{A}$.
  
  \begin{proof}
    The KKT condition of Eq.~{\eqref{eq:eq2quadprog}} indicates
    $\hat{X}_{\min} = \hat{q} - \hat{A} \Lambda_{\min}$ and $\hat{A}^{T} 
    \hat{X}_{\min} = \hat{b}$. This leads to $\hat{A}^{T}  ( \hat{q} - \hat{A}
    \Lambda_{\min} ) = \hat{b}$. Since $\hat{A}$ is full-rank,
    $\widehat{A^{}}^{T} \hat{A}$ is invertible. Hence we have
    \begin{equation}
      \Lambda_{\min} = ( \hat{A}^{T}  \hat{A} )^{-1}  ( \hat{A}^{T}  \hat{q} -
      \hat{b} ) .
    \end{equation}
    Plug-in $\hat{X}_{\min} = \hat{q} - \hat{A} \Lambda_{\min}$ yields
    Eq.~{\eqref{eq:exactsol}}.
  \end{proof}
\end{proposition}

Similarly, we also have
\begin{proposition}
  Suppose $\hat{X}^{\ast} ( \hat{Z} , \hat{Y} ) = \hat{N}  \hat{Z} + \hat{U} 
  \hat{D}  \hat{Y}$ is the solution of Eq.~{\eqref{eq:eqqpform}}, then
  \begin{equation}
    \label{eq:anavarsubspace} \hat{N} = ( \hat{C}^{+} )^{T} , \hspace{1em}
    \hat{U}  \hat{D} = (I- \hat{C}  \hat{C}^{+} )  \hat{D} ,
  \end{equation}
  where $\hat{C}^{+} = ( \hat{C}^{T}  \hat{C} )^{-1}  \hat{C}^{T}$ is
  pseudo-inverse of $\hat{C}$, and $\hat{U} =I- \hat{C}  \hat{C}^{+}$ is the
  orthogonal projector onto the kernel of $\hat{C}^{T}$
  {\cite{golub2012matrix}}.\\
  
\end{proposition}

Here we remark that in order for any subspace solution $\widehat{X^{}}^{\ast}
( \hat{Z} , \hat{Y} )$ to have a unique low-dimensional coordinate $( \hat{Z}
, \hat{Y} )$. We should require $\hat{C}$ and $\hat{D}$ to be linearly
independent. This equivalently means $\hat{U} \hat{D}$ is full rank.

\begin{theorem}[Projection on Variational Subspace]
  \label{projection} Assume columns of $\hat{C}$ and $\hat{D}$ are linearly
  independent. Given any $X \in \mathbb{R}^{n}$, 
  its closest point (under Euclidean distance) in a
  variational subspace $\hat{X}^{\ast} ( \hat{Z} , \hat{Y} )$ given by
  Eq.~{\eqref{eq:anavarsubspace}} is
  \begin{equation}
    \hat{Y} = ( \hat{U}  \hat{D} )^{+} X, \hspace{1em} \hat{Z} = \hat{C}^{T}
    X,
  \end{equation}
  and the closest point is
  \begin{equation}
    \hat{X}^{\ast} = ( \hat{U}  \hat{D} ( \hat{U}  \hat{D} )^{+}   \noplus +
    \hat{C} \hat{C}^{+} ) X,
  \end{equation}
  where $( \hat{U} \hat{D} )^{+} = ( \hat{D}^{T}  \hat{U}  \hat{D} )^{-1} \hat{D}^{T}  \hat{U}^{T}$ 
  and $\hat{U} =I- \hat{C} \hat{C}^{+}$.
  
  \begin{proof}
    If $\hat{U} \hat{D} v =0$ for some $v \in \mathbbm{R^{}}^{k}$, we have
    $\hat{D} v+ \hat{C} u=0$, where $u=C^{+} v$. Since $\hat{C}$ and $\hat{D}$
    are linearly independent, we have $u=0$ and $v=0$. Therefore, $\hat{U}
    \hat{D}$ is full-rank. Furthermore, \ $\hat{D}^{T} \hat{U} \hat{D} =
    \widehat{D^{}}^{T} \widehat{U^{}}^{T} \hat{U} \hat{D}$ is invertible. The
    closest point to $X$ is to minimize
    \[ \min  _{\hat{Z} , \hat{Y}} \|\hat{X}^{\ast} ( \hat{Z} , \hat{Y} )-X\|^{2} , \]
    whose partial gradient against $\hat{Z}$ and $\hat{Y}$ should be zero,
    i.e.,
    \[ \hat{N}^{T}  ( \hat{N}  \hat{Z} + \hat{U}  \hat{D}  \hat{Y} -X) =0 \]
    and
    \[ \hat{D}^{T}  \hat{U}^{T}  ( \hat{N}  \hat{Z} + \hat{U}  \hat{D}\hat{Y} -X) =0. \]
    Notice that $\hat{N}^{T}  \hat{U} =0$, 
    $\hat{N}^{T}  \hat{N} = (\hat{C}^{T}  \hat{C} )^{-1}$, $\hat{U}^{T}  \hat{U} = \hat{U}$. 
    Above equalities can be simplified to
    \[ ( \hat{C}^{T}  \hat{C} )^{-1}  \hat{Z} = \hat{N}^{T} X, \hspace{1em}
       \hat{D}^{T}  \hat{U}  \hat{D}  \hat{Y} = \hat{D}^{T}  \hat{U}^{T} X. \]
    Above equalities can be solved as
    \[ \hat{Z} = ( \hat{C}^{T}  \hat{C} )  \hat{N}^{T} X= \hat{C}^{T} X,
       \hspace{1em} \hat{Y} = ( \hat{D}^{T}  \hat{U}  \hat{D} )^{-1} 
       \hat{D}^{T}  \hat{U}^{T} X. \qedhere\]
  \end{proof}
\end{theorem}

Next, we are to derive the analytic subspace solution.
\begin{proposition}[Variational Subspace Solution]\label{vssol}
  Let $\hat{I} = \hat{U}  \hat{D} ( \hat{U}  \hat{D} )^{+}   \noplus +
  \hat{C} \hat{C}^{+}$ be orthogonal projector onto the subspace $\tmop{Span}
  (\hat C ) + \tmop{Span} (\hat D )$(\cite{yanai2011projection}, page 45), where $\hat{U} =I-
  \hat{C} \hat{C}^{+}$ the orthogonal projector onto the kernel of
  $\hat{C}^{T}$. Assuming $\hat{A}$ is full-rank, columns of $\hat{C}$ and
  $\hat{D}$ are linearly independent, and $( \hat{A}^{T}  \hat{I}  \hat{A}
  )^{-1}$exists, the variational subspace solution to
  Eq.~{\eqref{eq:eq2quadprog}} is
  \begin{equation}
    \label{eq:reducedsol} \hat{X}^{\ast}_{\min} = \left( \hat{I} - \hat{A}_{p} 
    \hat{A}_{p}^{+} \right)  \hat{q} + ( \hat{A}_{p}^{+} ) ^{T} \hat{b}
  \end{equation}
  where $\hat{A}_{p}  =  \hat{I} \hat{A}$ and 
  $\hat{A}_{p}^{+} = \left(\hat{A}^{T}_{p}  \hat{A}_{p} \right)^{-1}  \hat{A}_{p}^{T}$. 
  Note $\hat{I} -\hat{A}_{p}  \hat{A}_{p}^{+}$ is the projection matrix restricted in subspace
  $\tmop{Span} (\hat C ) + \tmop{Span} (\hat D )$ that map onto the kernel of $\hat{A}_{p}^{T}$.
  
  \begin{proof}
    First, we have $\hat{I}$ is symmetric, and $\hat{I} \hat{I}  =  \hat{I}$.
    Plug variational subspace $\hat{X}^{\ast} ( \hat{Z} , \hat{Y} )$ into
    Eq.~{\eqref{eq:eq2quadprog}}. From the KKT condition, we have
    \[ \left(\begin{array}{c}
         \hat{D}^{T}  \hat{U}^{T}\\
         \hat{N}^{T}
       \end{array}\right) \left[ \hat{N}  \hat{Z}_{\min} + \hat{U}  \hat{D} 
       \hat{Y}_{\min} + \hat{A} \Lambda^{\ast}_{\min} - \hat{q} \right] =0 \]
    and
    \begin{equation}
      \label{eq:constr} \hat{A}^{T}  \left[ \hat{N}  \hat{Z}_{\min} + \hat{U} 
      \hat{D}  \hat{Y}_{\min} \right] = \hat{b} .
    \end{equation}
    Similar to the derivation in Theorem~\ref{projection}, the former equality
    of KKT condition yields
    \[ \hat{Y}_{\min} = ( \hat{U}  \hat{D} )^{+}  ( \hat{q} - \hat{A}
       \Lambda^{\ast}_{\min} ) , \hspace{1em} \hat{Z}_{\min} = \hat{C}^{T}  (
       \hat{q} - \hat{A} \Lambda^{\ast}_{\min} ) , \]
    and
    \begin{equation}
      \label{eq:msol} \hat{X}^{\ast}_{\min} = \hat{I} ( \hat{q} - \hat{A}
      \Lambda^{\ast}_{\min} ) .
    \end{equation}
    Let $\widehat{X_{}}_{p}^{\ast} = \hat{q} - \hat{A} \Lambda^{\ast}_{\min}$,
    and combine Eq.~{\eqref{eq:msol}} with Eq.~{\eqref{eq:constr}}, we have
    $\hat{A}^{T}  \hat{I}  ( \hat{q} - \hat{A} \Lambda^{\ast}_{\min} ) = \hat{b}$. It gives
    \begin{equation}
      \label{eq:lambdamin} \Lambda^{\ast}_{\min} = ( \hat{A}^{T}  \hat{I} 
      \hat{A} )^{-1}  ( \hat{A}^{T}  \hat{I}  \hat{q} - \hat{b} ) .
    \end{equation}
    Plug Eq.~{\eqref{eq:lambdamin}} back to $\hat{X}^{\ast}_{\min}$
    (Eq.~{\eqref{eq:msol}}) yields the subspace solution
    Eq.~{\eqref{eq:reducedsol}}.
  \end{proof}
\end{proposition}

We are now ready to introduce the main result. 
Let $\|\cdot\|$ be the induced $L_{2}$ matrix
norm, which is its largest singular value.
\begin{theorem}[Approximation Error Bound of Variational Subspace Solution]
  \label{mainthm} Given the demand matrix $\hat{C}$ and $\hat{D}$ forming the
  subspace $\tmop{Span} ( C ) + \tmop{Span} ( D )$, where $\hat{U} =I- \hat{C}\hat{C}^{+}$, 
  and $\hat{I} = \hat{U}  \hat{D} ( \hat{U}  \hat{D} )^{+} + \hat{C} \hat{C}^{+}$. 
  The error between reduced solution
  $\hat{X}^{\ast}_{\min}$ to Eq.~{\eqref{eq:reducedsol}} and exact solution
  $\hat{X}_{\min}$ to Eq.~{\eqref{eq:exactsol}} has a following upper bound:
  Assuming $\| I- \hat{A}^{+} \hat{I} \hat{A} \|\le \rho <1$ 
  and $\tmop{cond}( \hat{A} ) = \| \hat{A}\| \| \hat{A}^{+} \| \le \omega <+ \infty$ for any
  $\hat{A}$ in the scope of optimization Eq.~{\eqref{eq:eq2quadprog}}, there
  exists constants $\beta_{1} >0$ and $\beta_{2} >0$, such that
  \begin{equation}
    \label{eq:bound} \| \hat{X}^{\ast}_{\min} - \hat{X}_{\min} \| \le \|
    \hat{I} \hat{q} - \hat{q} \| + \Delta \left( \hat{b} , \hat{q} , \hat{A}^{+} ;
    \beta_{1} , \beta_{2} \right)  \| \hat{I} \hat{A} - \hat{A} \| ,
  \end{equation}
  for any $\hat{q}$, $\hat{b}$, and $\hat{C}$, $\hat{D}$, $\hat{A}$, where
  \[ \Delta \left( \hat{b} , \hat{q} , \hat{A}^{+} ; \beta_{1} , \beta_{2} \right) =
     \beta_{1} \|\hat{b} \| \cdot  \| \hat{A}^{+} \|^{2} + \beta_{2} \| \hat{q} \|\cdot
     \| \hat{A}^{+} \| >0. \]
  In particular, if $\hat q=\hat D Y$ and $\hat A=\hat C  A_{c}$ for some $Y$ and $A_{c}$, then it
  must have $\hat{I} \hat{q} = \hat{q}$ and $\hat{I} \hat{A} = \hat{A}$, thus
  we know $\hat{X}^{\ast}_{\min} = \hat{X}_{\min}$.  
  \begin{proof} See Appendix~\ref{sec:mainproof}
  \end{proof}
\end{theorem}

Theorem~\ref{mainthm} bounds the approximation error between reduced solution
and exact solution by two terms. They are the norm of projections of $\hat{q}$
and $\hat{A}$ onto the intersection of kernel space of $\hat{D}^T$ and $\hat{C}^T$.
Finally, given the Prop.~\ref{equivalence}, we have
\begin{equation}\begin{split}
  \| X^{\ast}_{\min} -X_{\min} \|_{H} & = \| LX^{\ast}_{\min} -LX_{\min} \| \\
  &= \| LL^{+}  \hat{X}^{\ast}_{\min} -LL^{+}  \hat{X}_{\min} \| \\
  &\le \| LL^{+} \| \| \hat{X}^{\ast}_{\min} - \hat{X}_{\min} \| \\
  & \le \| \hat{X}^{\ast}_{\min} - \hat{X}_{\min} \| , 
\end{split}\end{equation}
where $X_{\min}$ is the solution to Eq.~{\eqref{eq:quadprog}} and
$X^{\ast}_{\min}$ is the corresponding variational reduced solution.

\section{Conclusions}\label{sec:conclusion}
In this paper, we presented variational subspace for reducing calculations in
minimizing quadratic functions subject to large-scale variables, and 
integrated it into an interactive modeling framework for mesh deformations. 
Variational subspace is an economical subspace driven by reduced constraint 
demands and optimization contexts. Based on it, we implemented an easy-to-use
mesh manipulator, which is efficent, robust in quality, 
intuitive to control, and extensible.

\noindent\textbf{Acknowledgment.}
The authors would like to thank anonymous reviewers in the past submission process
for their comments and suggestions. The authors also thank Prof. James Z. Wang for his
supports in the later stage of the work.


\bibliographystyle{acmsiggraph}

\bibliography{subspace}

\appendix
\setcounter{secnumdepth}{0}

\section{Proof of Theorem~\ref{mainthm}}\label{sec:mainproof}
We follow the notations used in section~\ref{sec:bound} to prove Theorem~\ref{mainthm}. 
We firstly have the following lemma. 

\begin{lemma}
  \label{lemmabound} Let $\hat{I} = \hat{U} \hat{D} ( \hat{U} \hat{D} )^{+}  
  \noplus + \hat{C} \hat{C}^{+}$ and \ $\hat{A}_{p}  =  \hat{I} \hat{A}$,
  where $\hat{U} =I- \hat{C} \hat{C}^{+}$. Assume $\| I- \hat{A}^{+} \hat{I}\hat{A} \| <1$ 
  (while we know $\| I- \hat{A}^{+} \hat{I} \hat{A} \| \le 1$,
  because $I- \hat{A}^{+} \hat{I} \hat{A}  $ is the projection matrix onto
  $\tmop{kernel} ( \{ \widehat{A^{}}^{+} [ \hat{U} \hat{D} , \hat{C} ] \}^{T})$ 
  and the equality holds if and only if $\tmop{rank} ( \widehat{A^{}}^{+} [
  \hat{U} \hat{D} , \hat{C} ] ) <m$), we have
  \begin{equation}
    \label{eq:lemmabound} \norm{ \hat{A} \left[ ( \hat{A}^{T}  \hat{A} )^{-1} - (
    \hat{A}^{T}_{p}  \hat{A}_{p} )^{-1} \right] \hat{A}^{T} } \le
    \dfrac{\tmop{cond} ( \hat{A} ) \| I- \hat{A}^{+} \hat{I} \hat{A} \|}{1- \|
    I- \hat{A}^{+} \hat{I} \hat{A} \|} ,
  \end{equation}
  where $\tmop{cond} ( \hat{A} ) = \| \hat{A} \| \| \hat{A}^{+} \|$ is the
  condition number of $\hat{A}$, and
  \begin{equation}
    \| \hat{A}^{+} - \hat{A}_{p}^{+} \| \le \dfrac{\norm{ \hat{A}^{+} }  \| I-
    \hat{A}^{+} \hat{I} \hat{A} \|}{1- \| I- \hat{A}^{+} \hat{I} \hat{A} \|} +
    \| \hat{A}^{+} |^{2} \|\hat{I} \hat{A} - \hat{A} \|.
  \end{equation}
  \begin{proof}
    We have the following expansion $\hat{A} \left[ ( \hat{A}^{T}  \hat{A} )^{-1} -
    ( \hat{A}^{T}_{p}  \hat{A}_{p} )^{-1} \right] \hat{A}^{T}  =  \hat{A}^{} ( I - 
    ( \hat{A}^{+} \hat{I} \hat{A}^{} )^{-1} ) \hat{A}^{+} = \hat{A} ( ( I-
    \hat{A}^{+} \hat{I} \hat{A} ) + ( I- \hat{A}^{+} \hat{I} \hat{A} )^{2} + (
    I- \hat{A}^{+} \hat{I} \hat{A} )^{3} + \ldots ) \hat{A}^{+}$. 
    Therefore, we have 
    \begin{equation}
      \begin{split}
        & \norm{\hat{A} \left[ ( \hat{A}^{T}  \hat{A} )^{-1} - ( \hat{A}^{T}_{p} 
        \hat{A}_{p} )^{-1} \right] \hat{A}^{T}}  \\ 
      & \le \| \hat{A} \|   ( \| I-
      \hat{A}^{+} \hat{I} \hat{A} \| + \| I- \hat{A}^{+} \hat{I} \hat{A} \|^{2}
      + \| I- \hat{A}^{+} \hat{I} \hat{A} \|^{3} + \ldots ) \| \hat{A}^{+} \|\\
      & \le \dfrac{\tmop{cond} ( \hat{A} ) \| I- \hat{A}^{+} \hat{I} \hat{A}
        \|}{1- \| I- \hat{A}^{+} \hat{I} \hat{A} \|}.
      \end{split}
    \end{equation}
    Similarly, we have \begin{equation}\begin{split}
    \|\hat{A}^{+} - \hat{A}_{p}^{+}\| 
    &= \| [ ( \hat{A}^{T}  \hat{A} )^{-1} - (
    \hat{A}^{T}_{p}  \hat{A}_{p} )^{-1} ] \hat{A}^{T} \hat{I} \| + \|
    \hat{A}^{+} - \hat{A}^{+} \hat{I} \| \\ 
    & \le \dfrac{\| \hat{A}^{+} \|   \| I-
    \hat{A}^{+} \hat{I} \hat{A} \|}{1- \| I- \hat{A}^{+} \hat{I} \hat{A} \|} +
    \| \hat{A}^{+} \|^{2} \| \hat{I} \hat{A} - \hat{A} \|. \qedhere
    \end{split}
    \end{equation}
  \end{proof}
\end{lemma}

\begin{corollary}
  \label{lemmabound2} Since $\| I- \hat{A}^{+} \hat{I} \hat{A} \| =
  \| \hat{A}^{+} \hat{A} - \hat{A}^{+} \hat{I} \hat{A} \| \le \| \hat{A}^{+}
  \|   \| \hat{I} \hat{A} - \hat{A} \|$, we have 
  \[\norm{ \hat{A} \left[ ( \hat{A}^{T} 
  \hat{A} )^{-1} - ( \hat{A}^{T}_{p}  \hat{A}_{p} )^{-1} \right] \hat{A}^{T} } \le 
  \dfrac{\tmop{cond} ( \hat{A} ) \| \hat{A}^{+} \|}{1- \| I- \hat{A}^{+}
  \hat{I} \hat{A} \|} \| \hat{I} \hat{A} - \hat{A} \|,\] 
  and \[\| \hat{A}^{+} - \hat{A}_{p}^{+} \| 
  \le \dfrac{\| \hat{A}^{+} \|^{2} ( 2- \| I- \hat{A}^{+}
  \hat{I} \hat{A} \| )}{1- \| I- \hat{A}^{+} \hat{I} \hat{A} \|} \| \hat{I}
  \hat{A} - \hat{A} \|.\]
\end{corollary}

Here we prove the main result
\begin{proof}
  Based on Prop~\ref{exactsol} and Prof~\ref{vssol}.
  We can decompose the error term into two parts:
  \begin{equation}
    \| \hat{X}^{\ast}_{\min} - \hat{X}_{\min} \| \le
    \| ( \hat{A}^{+} )^{T} \hat{b} - ( \hat{A}_{p}^{+} ) \hat{b}\| + 
    \|( \hat{I} - \hat{A}_{p}  \hat{A}_{p}^{+} )  \hat{q} - ( I- \hat{A} 
    \hat{A}^{+} )  \hat{q}\|.
  \end{equation}
  
  On one hand, based on Corollary~\ref{lemmabound2}, there exists a constant $\beta_{1} = \frac{2-
    \rho}{1- \rho} >0$, such that
  \begin{equation}
    \begin{split}
      \| ( \hat{A}^{+} )^{T} \hat{b} - ( \hat{A}_{p}^{+} ) \hat{b} \| & \le  
      \| \hat{A}^{+} - \hat{A}_{p}^{+} \|   \| \hat{b} \| \\
      & \le \beta_{1} \|
      \hat{b} \|   \| \hat{A}^{+} \|^{2} \| \hat{I} \hat{A} - \hat{A} \|
    \end{split} \label{eq:mainpart1}
  \end{equation}
  On the other hand, from Corollary~\ref{lemmabound2}, we also have
  \begin{equation}\label{eq:mainpart2}
    \begin{split}
      &\| ( \hat{I} - \hat{A}_{p}  \hat{A}_{p}^{+} )  \hat{q} - ( I- \hat{A} 
      \hat{A}^{+} )  \hat{q} \| \\
      & \le \| ( \hat{I} - \hat{A}_{p}  \hat{A}_{p}^{+} )  \hat{q} - ( I-
      \hat{A}  \hat{A}^{+} ) \hat{I}  \hat{q} \| + \| I- \hat{A} 
      \hat{A}^{+} \| \| \hat{I} \hat{q} - \hat{q} \|\\
      & \le \| ( \hat{I} - \hat{A}_{p}  \hat{A}_{p}^{+} )  \hat{q} - \hat{I} (
      I- \hat{A}  \hat{A}^{+} ) \hat{I}  \hat{q} \| + \| \hat{A}^{+} \hat{I}
      \hat{q} \| \| \hat{I} \hat{A} - \hat{A} \| \\
      &\quad + \| I- \hat{A}  \hat{A}^{+} \| \| \hat{I} \hat{q} - \hat{q} \|\\
      & \le \left( 1+ \frac{\tmop{cond} ( \hat{A} )}{1- \rho} \right)   \|
      \hat{q} \|   \| \hat{A}^{+} \| \| \hat{I} \hat{A} - \hat{A} \|  + \|
      \hat{I} \hat{q} - \hat{q} \| ,
    \end{split}
  \end{equation}
  where $\beta_{2} =1+ \dfrac{\omega}{1- \rho} >0$. Combining
  Eq.~{\eqref{eq:mainpart1}} and Eq.~{\eqref{eq:mainpart2}},
  Eq.~{\eqref{eq:bound}} is held.
\end{proof}

\section{Implementation Details}\label{sec:implementation}
In companion to our proposed algorithm, other algorithm details less relevant to 
variational subspace is provided in this section, which we follow the notations
used in section~\ref{sec:ourapproach}.



\begin{figure}[t]
    \includegraphics[width =0.45\textwidth]{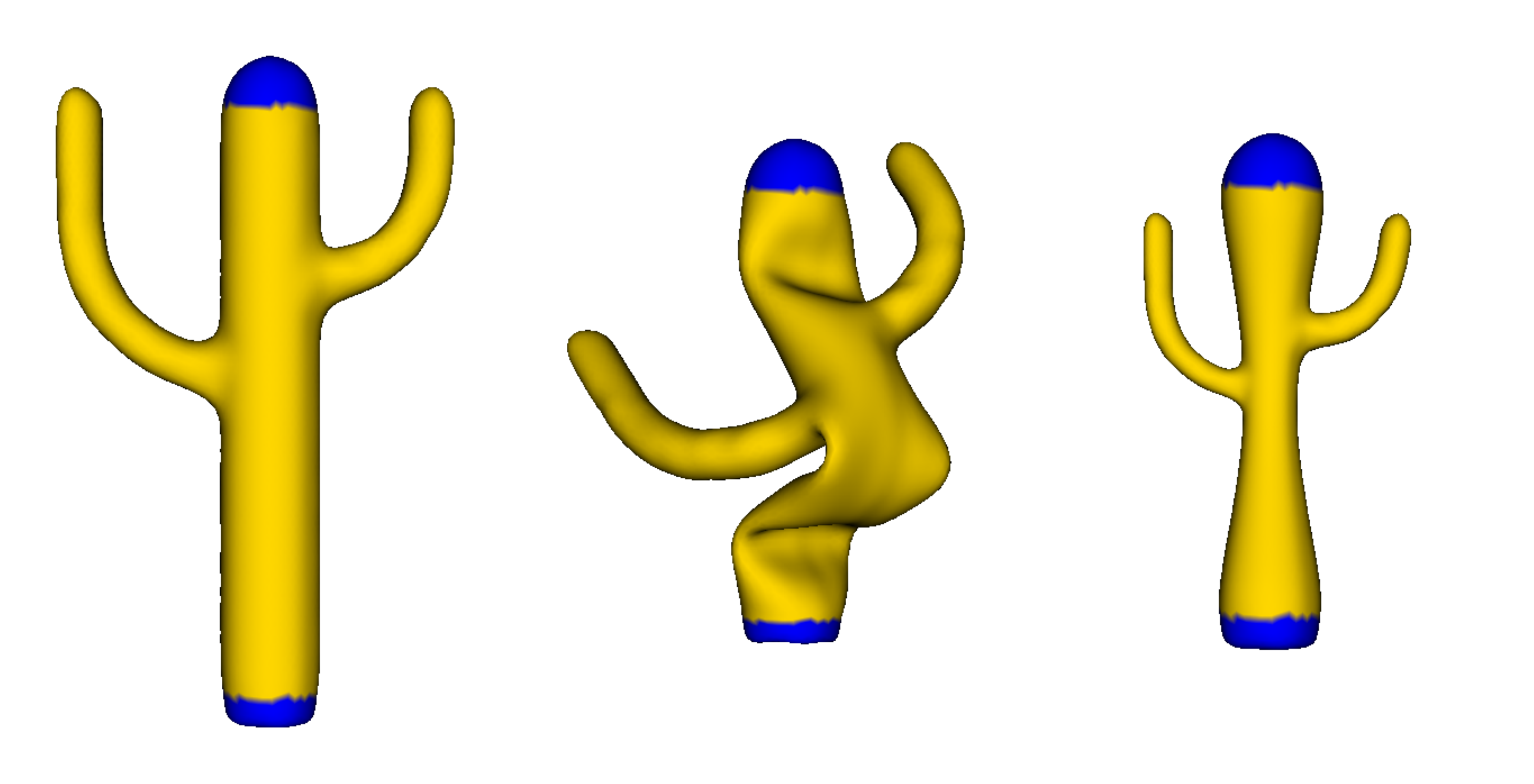}
    \caption{Left: Original model; Middle: Out-of-shape distortion in ARAP surface modeling; Right: Our method deforms rest-pose part via shrinking.}
    \label{fig:arapbad} 
\end{figure}

\begin{figure}[t]
    \includegraphics[width=0.45\textwidth]{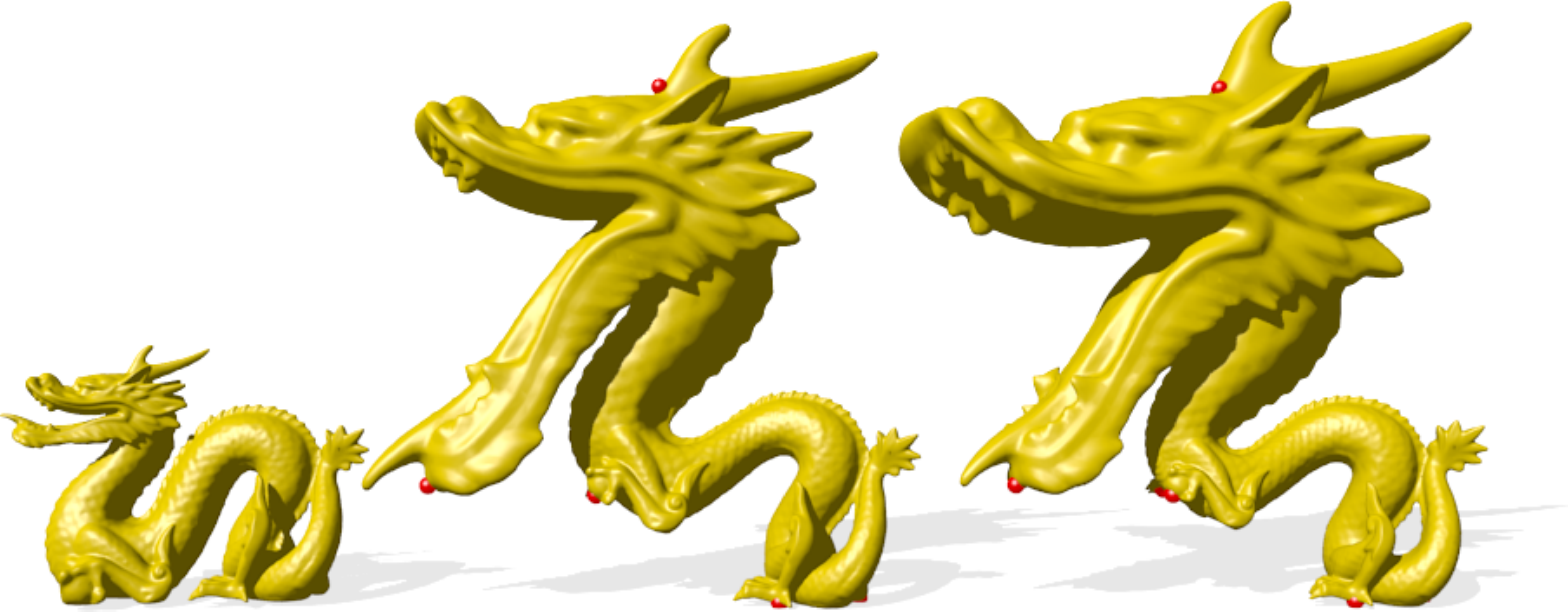} 
    \caption{Difference between deformations with conformal factors (right) and without (middle).
      Seven constraint points are marked as red balls.}
    \label{fig:conformal}
\end{figure}

\subsection{Global rotation adaption}
In our implementation, we also introduce global rotation $\mathbf r_0$ to
diminish the approximation error of local rotation matrix $\mathbf r_k$
incurred by piece-wise linear form (see equation \eqref{eq:rolinear}).
It is fitted again by a single SVD in each frame.
Then we update the reduced model (Phase 1 and 2) under updated frame coordinates,
i.e., multiply the inverse rotation $\mathbf r_0^{T}$ to $P_{\mbox{\scriptsize eq}}$ and $S$.
Meanwhile during mesh reconstruction (based on Eq. \eqref{eq:ARAPsubspace}),
we should also multiply rotation $\mathbf r_0$ to vertices $V'$,
so as to display them in the original frame.

\subsection{Affine Patches}
In deformable modeling, the user usually would like to constrain certain patches on the mesh
to be rigid or fixed, or more generally affine. Our framework can be accompanied by
those requirements in pre-computation. Vertex positions $\mathbf V'$ on the deformed mesh
can be linearly expressed in terms of deformable
vertices $\mathbf V'_0$ and patch-wise transformation $\mathbf t_i,\mathbf d_i$,
i.e., (under a permutation)
\begin{equation}
  \mathbf V'= [\mathbf V'_0; \mathbf V_1 \mathbf t_1 + \mathbf d_1; \ldots;
  \mathbf V_s \mathbf t_s + \mathbf d_s]\;,
\end{equation}
where $\mathbf V'_0$ are deformable vertices, $\mathbf V_1,\ldots,\mathbf V_s$
are $s$ affine patches on the original mesh with prescribed transformation matrices
$\mathbf t = [\mathbf t_1,\ldots,\mathbf t_s]$, and
displacements $\mathbf d = [\mathbf d_1,\ldots,\mathbf d_s]$.
Under this representation, the first stage problem
is reformulated accordingly such that the variational subspace is solved for
variables of the \textit{de facto} control layer $[\mathbf V'_0, \mathbf t, \mathbf d, \mathbf Q]$,
instead of for $[\mathbf V', \mathbf Q]$ (see equations \eqref{eq:KKTsubspace} and \eqref{eq:ARAPsubspace}). For simplicity, each affine patch accompanies a single rotational proxy
and a single linear proxy.

To improve the numerical stability in case one would like to constrain more than one vertex
on a single affine patch (e.g., constrain four in rigid motion), we in addition append 
corresponding linear proxies for each variable of $\mathbf t$. Therefore, the total degree of linear
proxies is $3m + 9s$.

\subsection{Conformal-like Deformations}

We extend our model to conformal-like deformations in this section,
by introducing scaling factors $\mathbf s_i$ for each rotational proxy.
Instead of restricting $\mathbf s_i\in SO(3)$, we permit
$\frac{\mathbf s_i}{\norm{\mathbf s_i}_2} \in SO(3)$,
where $\norm{\mathbf s_i}_2 \in [1/\psi, \psi]$, for some constant $\psi>1.$

Thus we can write $\mathbf s_i = \psi_i \mathbf t_i$,
where $\psi_i\in [1/\psi, \psi]$ and $\mathbf t_i \in SO(3)$.
The updating routine also contains two phases in correspondence,
of which the first phase is identical to former.
For the second phase, we reformulate as follows.

Similar to the previous Phase 2, we are again to fit the consistent local frame $\mathbf s_i$
by optimizing the simplified energy
\begin{equation}
  \begin{array}{rl}
  \mathcal E(\mathbf V'^{(i)},\mathbf S) = & \mbox{constant} - [V'^{(i)};\mathbf 0]^TM^T\left(T\circ(\Psi \otimes \mathbf 1_{9\times 1})\right)\\
  &+ \frac 12 \sum\limits_{k=1}^r\sum\limits_{(i,j)\in \mathcal G_k} c_{ijk}\norm{\mathbf v_i-\mathbf v_j}^2 \psi_{i_k}^2\;,
\end{array}
\end{equation}
where $T_{9d\times 1}$ is the vectorization of $(\mathbf t_i)$, $\Psi = [\psi_1,\ldots,\psi_d]$
and $S = T\circ(\Psi \otimes \mathbf 1_{9\times 1})$. To solve for $S$,
we use two steps: first, we fix $\Psi$ and optimize $T$, which is
exactly the same as discussed. It is required to compute
$(M_N X^{(i)} + M_US^{(i)})$ and perform singular value decomposition.
Second, we fix $T$ and compute partial gradient w.r.t. $\Psi$
\begin{equation}\begin{array}{rcl}
    \dfrac{\partial \mathcal E}{\partial \Psi} &=&  -(I_{d\times d}\otimes \mathbf 1_{1\times 9})T\circ M [V'^{(i)};\mathbf 0] + C\circ \Psi\\
    &=& -(I_{d\times d}\otimes \mathbf 1_{1\times 9})T\circ (M_N X^{(i)} + M_US^{(i)})  + C\circ \Psi\;,
  \end{array}
\end{equation}
where $C_{d\times 1}$ is pre-computed and $(M_N X^{(i)} + M_US^{(i)})$ is computed
in the former step. Hence by setting $\dfrac{\partial \mathcal E}{\partial \Psi} = 0$,
$\Psi$ is computed. Fig.~\ref{fig:conformal} illustrates the
difference between deformations with conformal factors and without.

\end{document}